\documentclass[aps,twocolumn,superscriptaddress,floatfix,letterpaper]{revtex4-1}
\usepackage[inline]{enumitem}
\usepackage{graphicx}

\pdfoutput=1

\newcommand{\Id}{\operatorname{Id}}

\begin{document}

\begin{titlepage}

\title{
Realizing Majorana zero mode
with a stripe of \emph{ionized} and \emph{isolated} magnetic atoms 
on a \emph{layered} superconductor
}

\author{Wenjie Ji}
\affiliation{Department of Physics, Massachusetts Institute of
Technology, Cambridge, Massachusetts 02139, USA}

\author{Xiao-Gang Wen}
\affiliation{Department of Physics, Massachusetts Institute of
Technology, Cambridge, Massachusetts 02139, USA}

\begin{abstract} 
It has been proposed that a line junction between spiral magnet and
superconductor or between ferromagnet and superconductor with Rashba
spin-orbital coupling can produce Majorana zero mode (MZM) at the ends of the
line.  However, a strong magnetic exchange coupling between the magnetic atoms
and the superconductor (about half of the bandwidth of the superconductor)  is
needed to obtain MZM.  Here, we design devices to reduce the needed magnetic
exchange coupling.  In the first proposal, we cover a very narrow $s$-wave
superconducting wire formed by a monolayer film or a layered material (such as
FeSe or FeSe monolayer on SrTiO$_3$) with \emph{isolated} magnetic atoms.  In
our second proposal, we place a line of \emph{isolated} magnetic \emph{ions} on
a monolayer superconductor or a \emph{layered} superconductor (such as FeSe).
We show that, in the above devices, a spiral magnetic order will develop
spontaneously to produce a 1D $p$-wave topological superconductor with a
sizable gap (above 1/2 of the parent superconducting gap), even for a weak
magnetic exchange coupling (less than 1/10 of the bandwidth).  The topological
superconductor has MZM at ends of the wire.  

\end{abstract}

\pacs{}

\maketitle

\end{titlepage}

\noindent \textbf{Introduction:} 
Topological order \cite{Wtoprev} describes gapped phases of quantum matter at
zero temperature that are robust against any perturbations (include those that
break all the symmetries).  For fermion systems, the simplest topological order
is the so-called invertible topological order (iTO),\cite{KW1458,F1478} such as
2D integer quantum Hall states \cite{KDP8094}, as well as 1D $p$-wave
\cite{JM7114,K0131} and  2D $p+\ii p$-wave topological superconductors
(TSC).\cite{SMF9945,RG0067}  The iTOs have no non-trivial bulk topological
excitations.  Their non-trivialness is reflected by their boundary states. 

For example, the 1D $p$-wave TSC chain of length $L$ has a 2-fold
\emph{topological degeneracy} in \emph{many-body} energy levels. (A topological
degeneracy is a non-exact degeneracy with splitting $\del \sim \ee^{-L/\xi} \to
0$ as sample size $L\to \infty$.  Such a non-exact degeneracy is robust against
any perturbations that can break any symmetry.\cite{WN9077}) The topological
degeneracy can be used as logical qubits in topological quantum
computing.\cite{K032}  Since the 2-fold topological degeneracy comes from the
two ends of the chain (there is no degeneracy for a $p$-wave superconducting
ring), each end carries degrees of freedom of one-half of a qubit: the Hilbert
space for degrees of freedom at one end has a \emph{non-integer} dimension
$\sqrt 2$ (which is called the quantum dimension)! The emergence of non-integer
degrees of freedom is a unique character of topological
orders\cite{W9102,MR9162}.

If the fermions in the chain are non-interacting, then the 2-fold topological
degeneracy of the chain can be explained by the two Majorana zero modes (MZM)
\cite{RG0067,I0168} at the two ends of the chain.\cite{JM7114,K0131} In this
case, the non-integer degrees of freedom correspond to a Majorana zero mode.
However, for interacting fermions, there are no single particle levels and no
MZMs.  In this case, we cannot regard the 2-fold topological degeneracy in
terms of MZMs.

\begin{figure*}[tb] 
\includegraphics[height=1.6in]{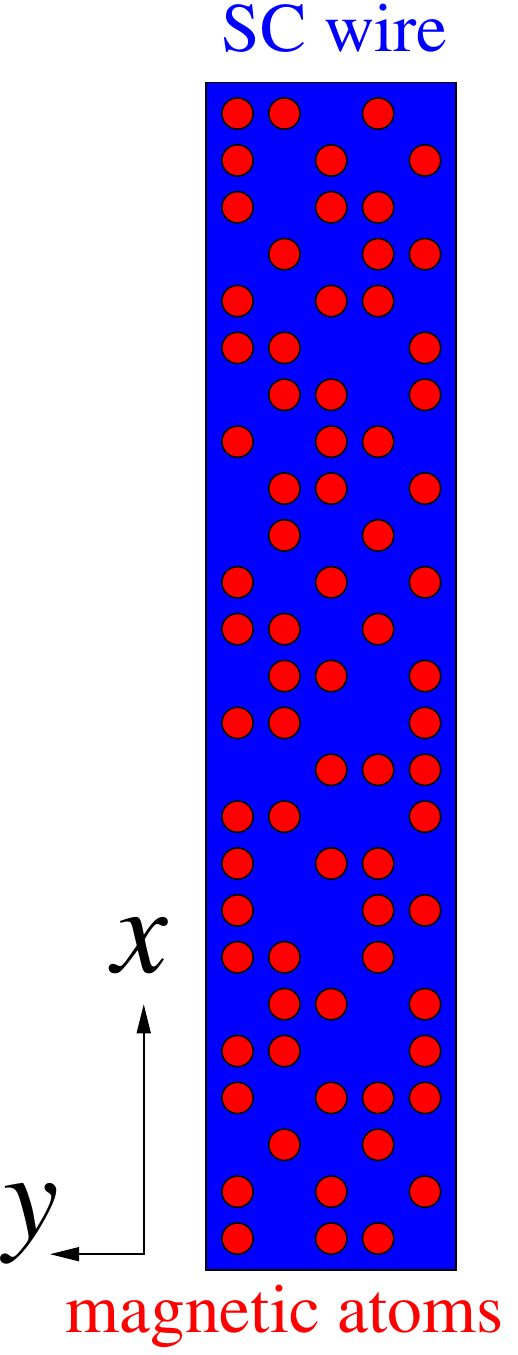}~~~~~~ 
\includegraphics[height=1.6in]{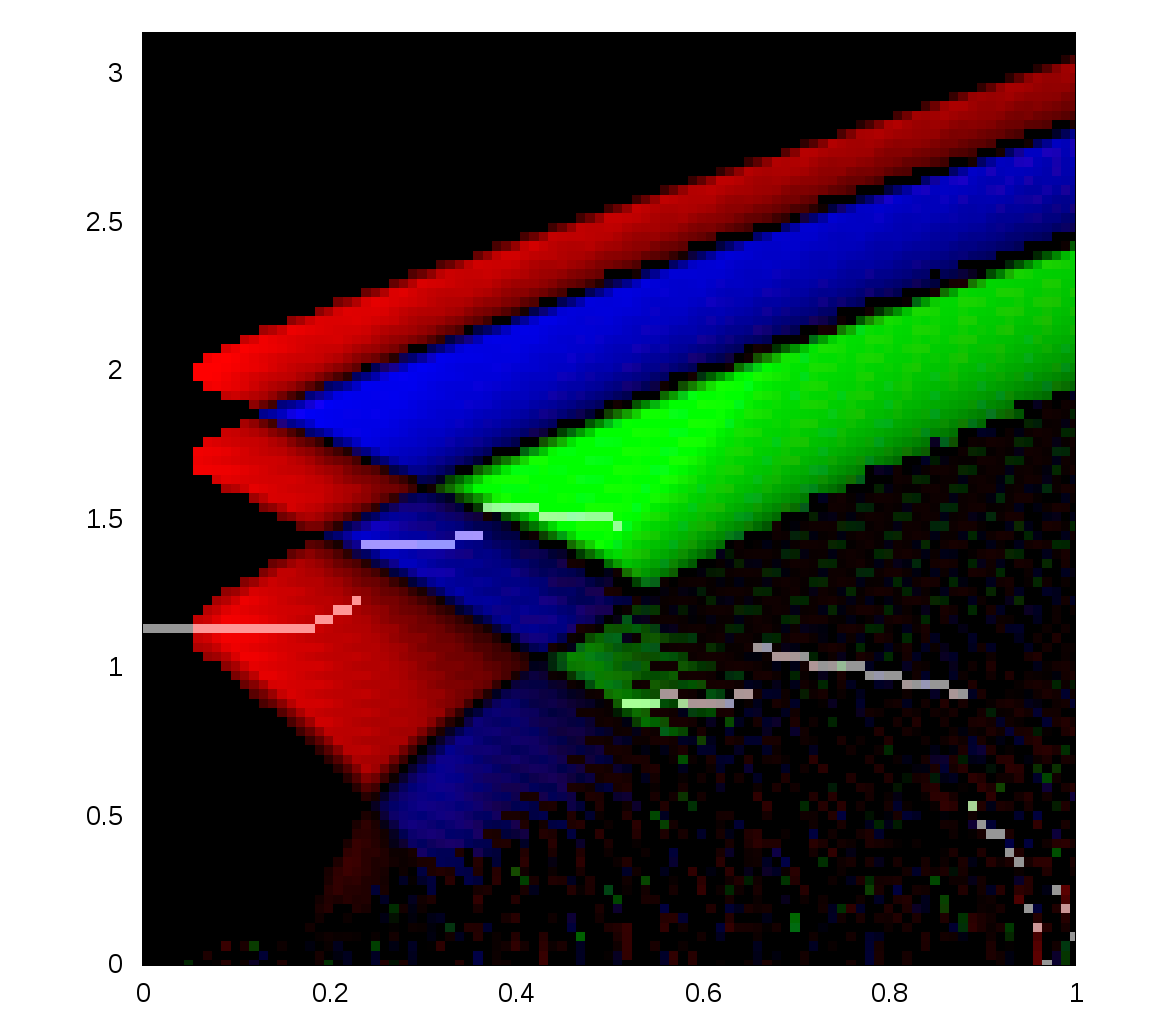}~ 
\includegraphics[height=1.6in]{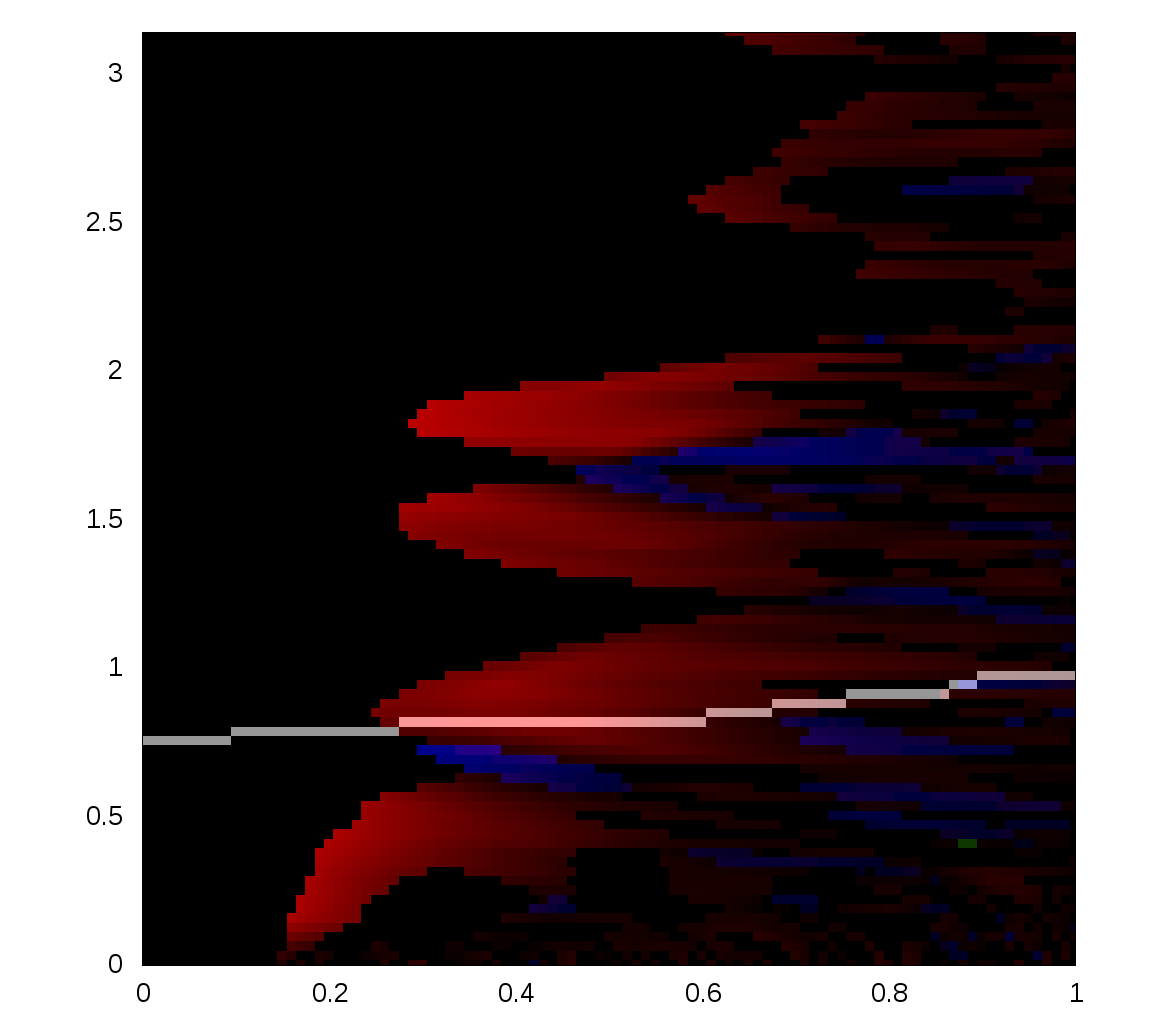}~ 
\includegraphics[height=1.6in]{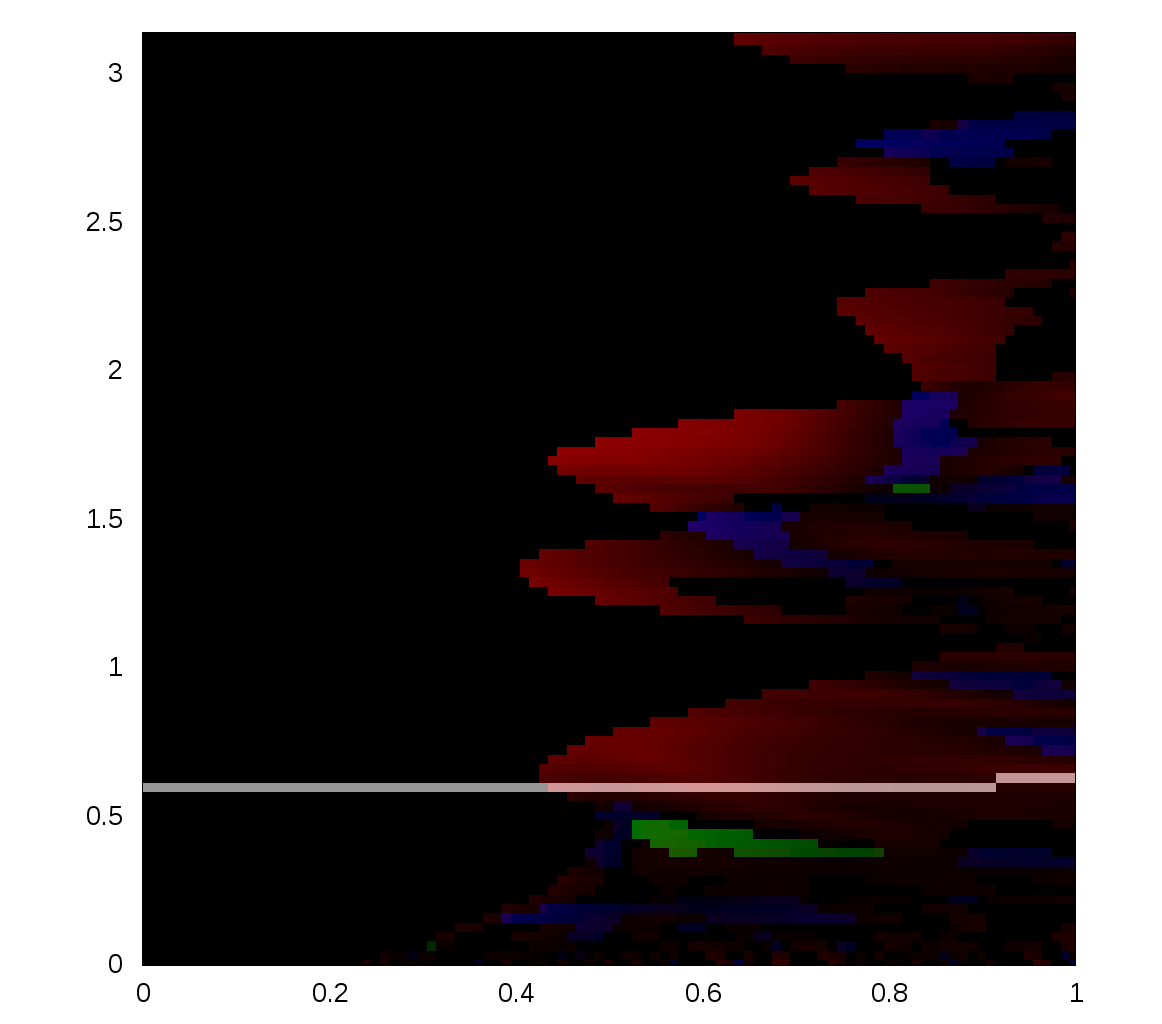}~
\includegraphics[height=1.6in]{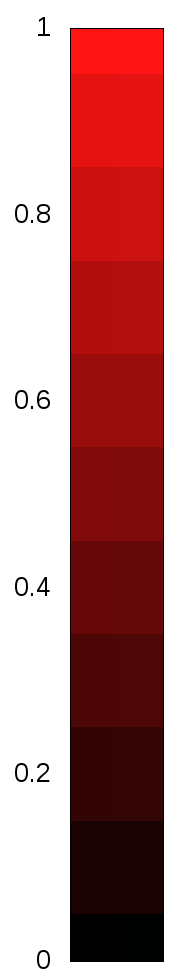} 
\\
\hfill
\hskip 0in (a)
\hskip 1.3in (b)
\hskip 1.7in (c)
\hskip 1.7in (d) ~~~~~~~~~~~~~~~~~~~~(e) ~~~\\
\caption{ 
The phase diagram of the first device: (a) isolated magnetic atoms on a layered
SC wire of size $100a\times 10a$, where $a$ is the lattice spacing.  The horizontal axis is the MEC $J$ and the vertical axis is the
spiral wave-vector $k^M_x a$.  The red area is the TSC phase with one MZM at an
end of the wire.  The green area is the TSC phase with three MZMs.  The blue
area is the SPT phase with two MZMs.  
The white line marks the spiral wave-vector
$k^M_x$ that corresponds to the ground state.  (b) 1.0 magnetic atoms per site
(uniform distribution) with $V=0$.  (c) 0.3 magnetic atoms per site with
$V=0.5$.  (d) 0.3 magnetic atoms per site with $V=1.0$.  
(e) The brightness of the color represents
the gap $\Del_\text{TSC}$ of the TSC, with the full brightness corresponds to 
$\Del_\text{TSC}/\Del=1$,  
where $\Del$ is the parent superconducting gap. 
}
\label{first} 
\end{figure*}

\begin{figure}[tb] 
\centering 
\includegraphics[height=0.4in]{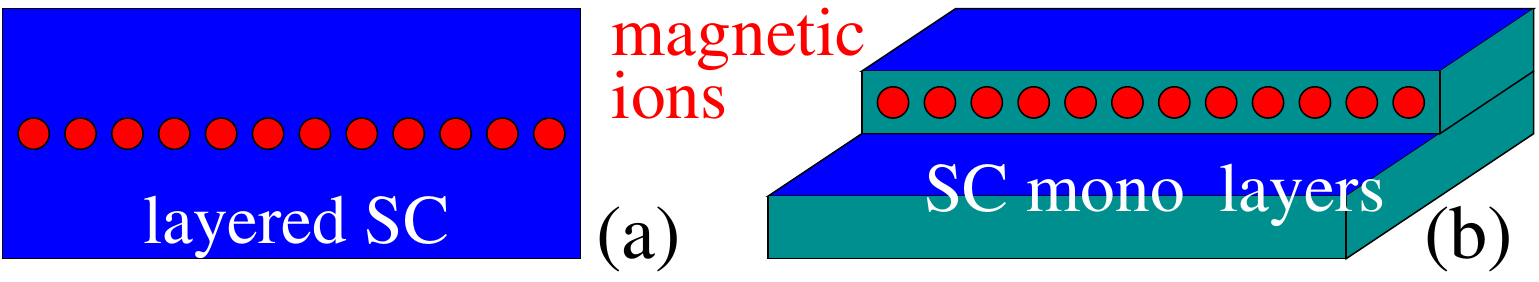} 
\caption{ 
(a) The second device made of isolated magnetic ions on the surface of a
layered SC. (b) Magnetic ions on a step edge of a SC layer.  The MZMs are at
the ends of the magnetic chain.
}
\label{step} 
\end{figure}

There are many proposals to realize the 1D TSC.  One class is to use a line
junction between an $s$-wave superconductor and a metallic quantum wire with
spin-orbital coupling, in an external magnetic field or in the presence of
magnetic moments in the  metallic
wire.\cite{SD09072239,OV10031145,PL12012176,KL12077322,KD14017048} In this
class of devices, the active electrons in the metallic wire may form a TSC due
to the proximity effect of the superconductor.  Another class is to use a line
junction between a superconductor and an insulating ferromagnet with Rashba
spin-orbital coupling, or between a superconductor and an insulating spiral
magnet.\cite{CB11080419,NY13036363,KL13071442,BS13072431,VF13072279,RF14065222,DRT1412,SP150907399,XA151008308,CP160708190}
In the second class of devices, Yu-Shiba-Rusinov states
\cite{Y6575,S6835,R6985} in the superconductor, induced by the magnetic
exchange coupling (MEC) between the magnetic insulator and the superconductor,
may form a TSC.  
However, for the second class of proposals, a strong MEC about
the bandwidth is needed to obtain 1D TSC phase (see Fig.  \ref{SCbSMI}),
even though the superconducting (SC) gap is much smaller.

In \Ref{KL13071442,BS13072431,VF13072279,RF14065222,SP150907399,CP160708190},
devices formed by a \emph{uniform} and \emph{insulating} line of magnetic atoms
on a bulk $s$-wave superconductor is proposed.  It was shown that a spiral
magnetic order of the magnetic atoms will develop automatically to produce an
1D TSC. A strong MEC (about the bandwidth) is still needed to realize the 1D
TSC in such a design (see Supplementary Materials
Fig.  \ref{thBkO}:Left).  In this paper, we design
devices to produce the 1D TSC with a weak MEC, by using SC wire or by using
magnetic atoms that are ionized on the superconductor.

\noindent \textbf{Device design:} 
In the first device (see Fig. \ref{first}a), we use a \emph{layered
SC wire} to realize 1D TSC.
The layered SC wire is formed by an $s$-wave  monolayer
thin film or a layered $s$-wave SC material.  
The width of the wire should be less than the superconducting coherent length
(smaller is better). We also cover the wire with \emph{isolated} magnetic atoms
which can have a uniform or a random distribution (the more uniform the
better).  In this case, we find that, even for a weak MEC (about 1/4 of the
Fermi energy or 1/16 of bandwidth), a spiral magnetic order of the magnetic
atoms will develop automatically to produce an 1D TSC with a sizable gap (about
1/2 of the parent superconducting gap).  If the  magnetic atoms have a perfect
uniform distribution (\ie form a periodic lattice), the required MEC is even
smaller (of order of the interlayer hopping amplitude or the gap of the parent
superconductor), and the gap of the induced TSC is close to that of the parent
superconductor (see Fig.  \ref{first}b).  

\begin{figure*}[tb] 
\centering 
\includegraphics[height=1.5in]{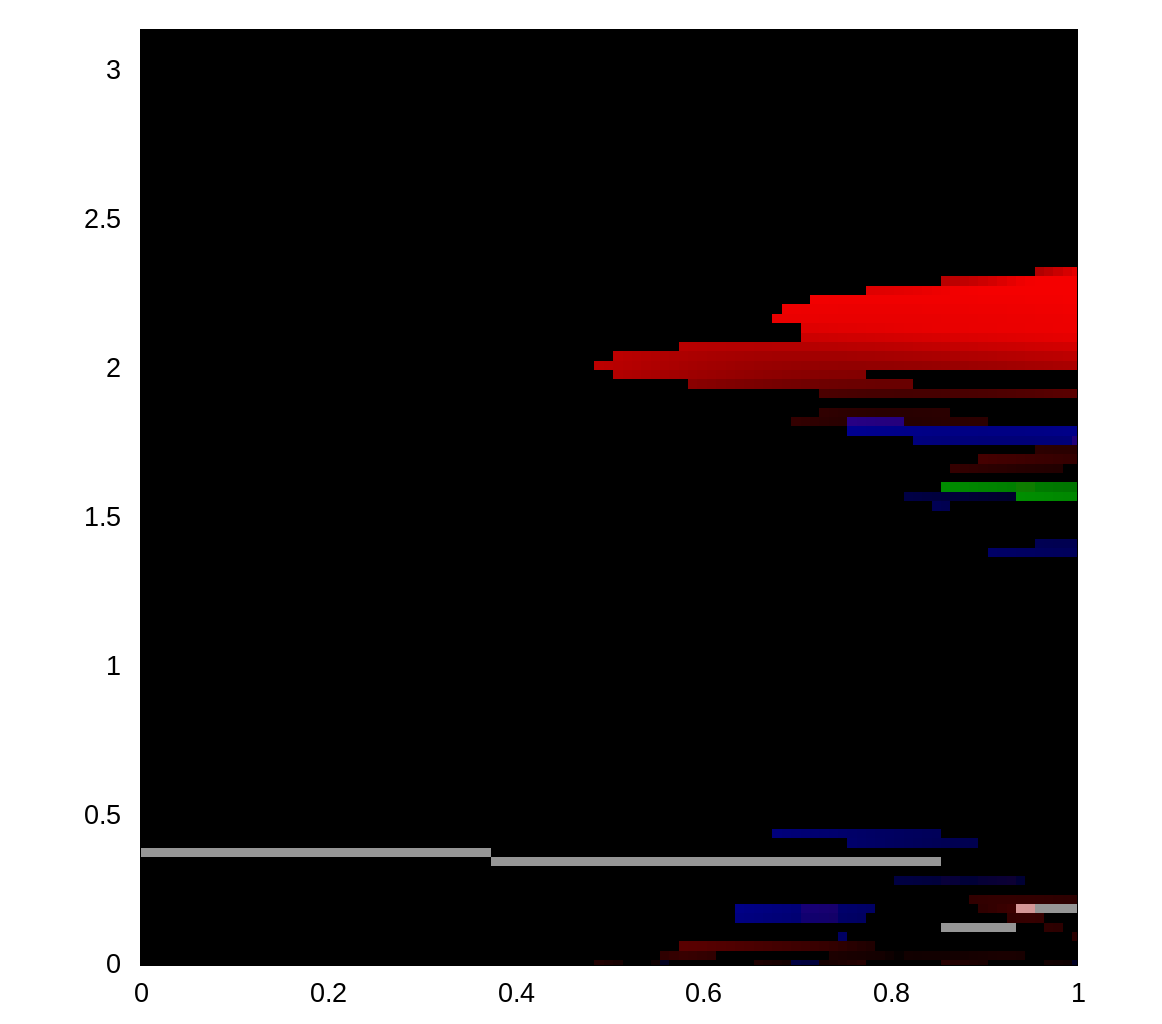} 
\includegraphics[height=1.5in]{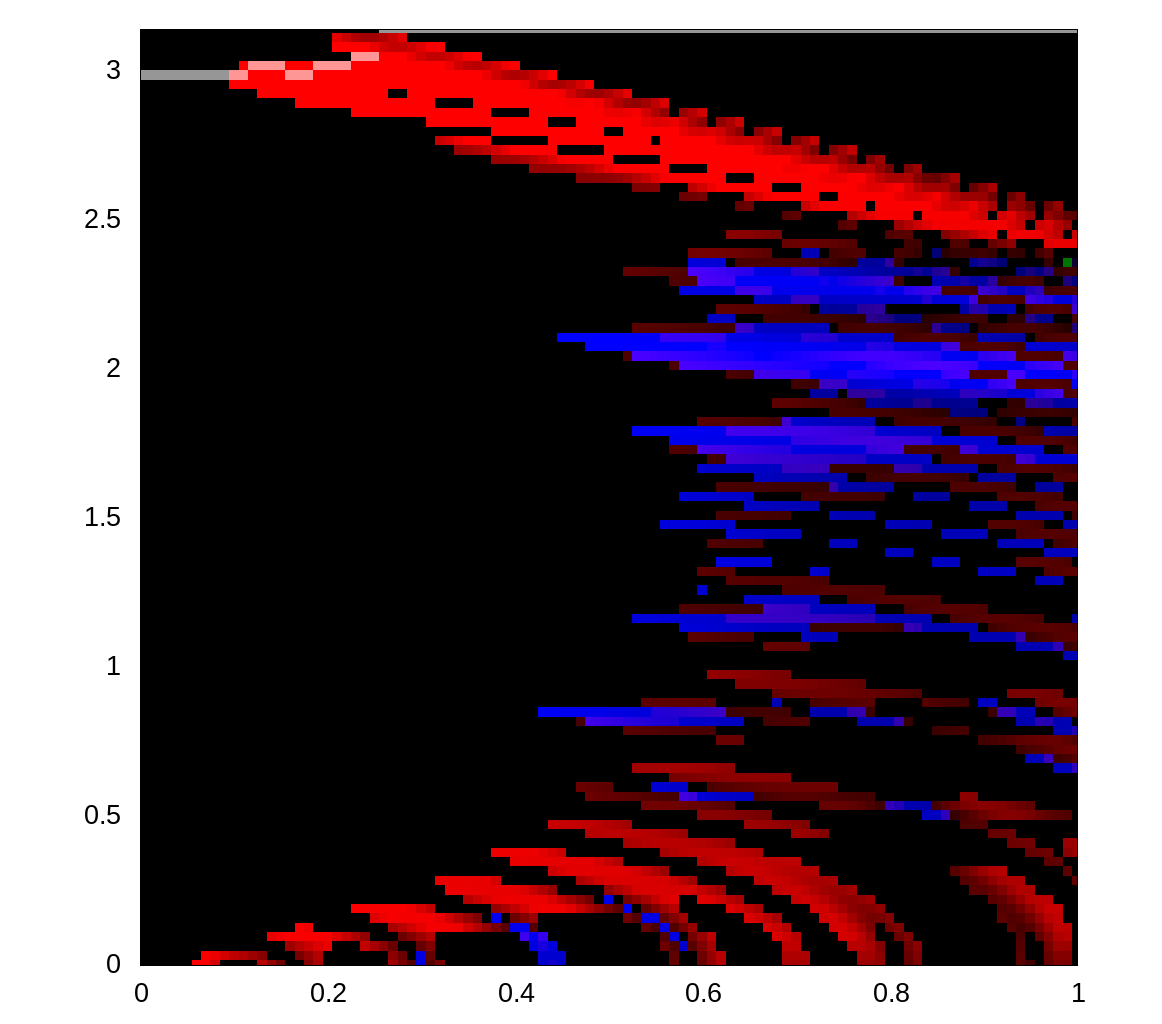} 
\includegraphics[height=1.5in]{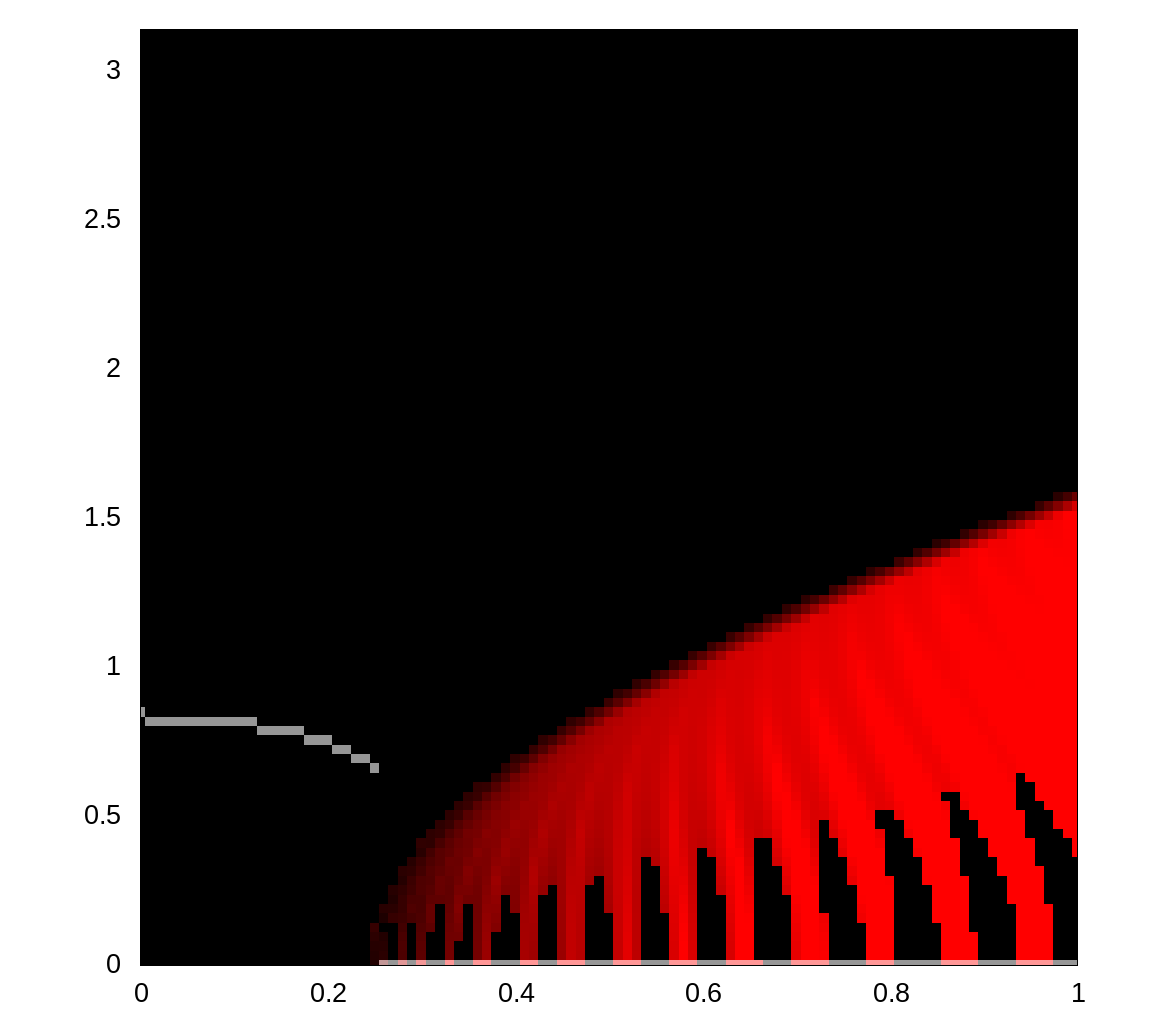} 
\includegraphics[height=1.5in]{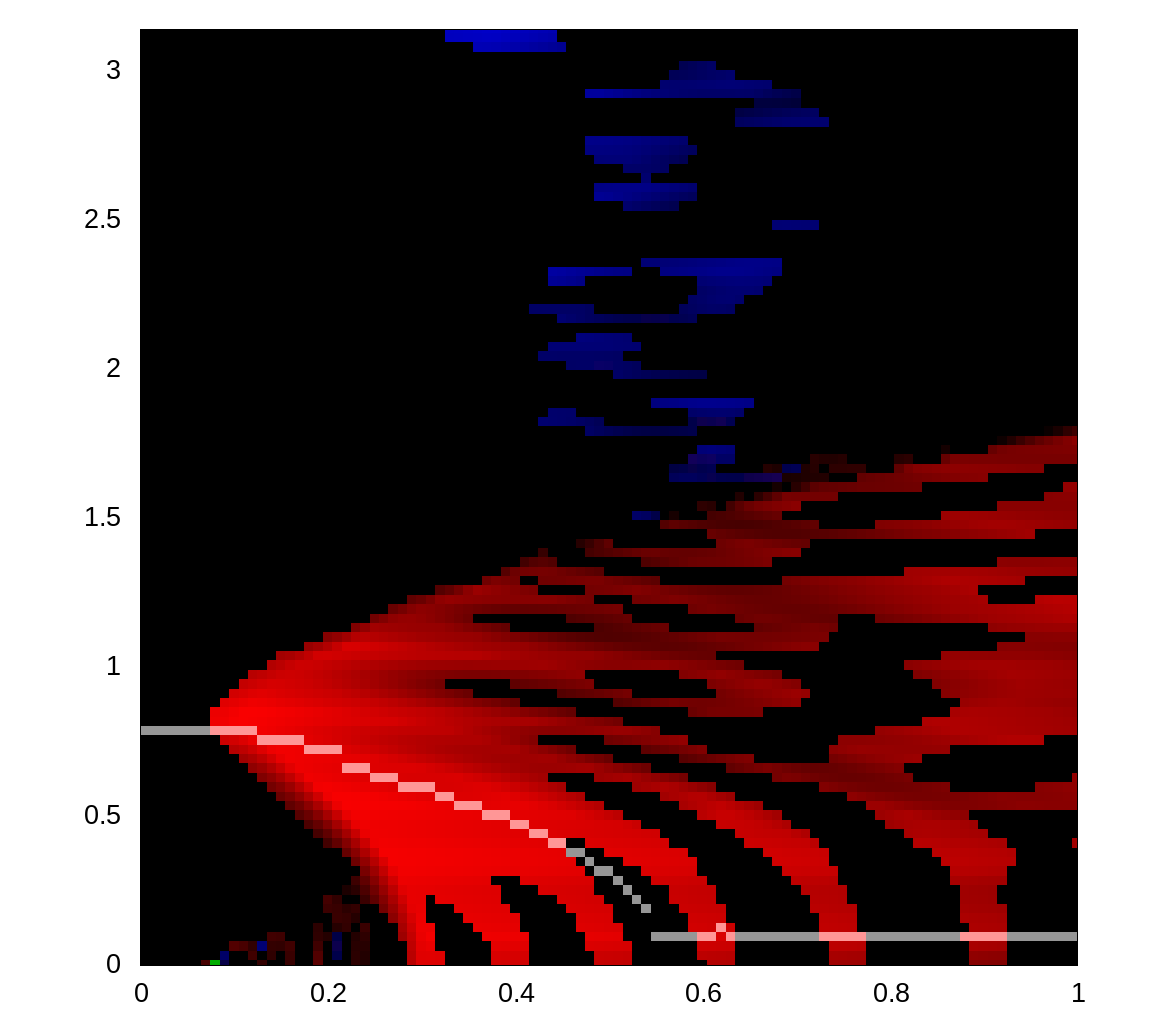} 
\hskip 0.9in (a)
\hskip 1.55in (b)
\hskip 1.55in (c)
\hskip 1.55in (d)\\
\includegraphics[height=1.5in]{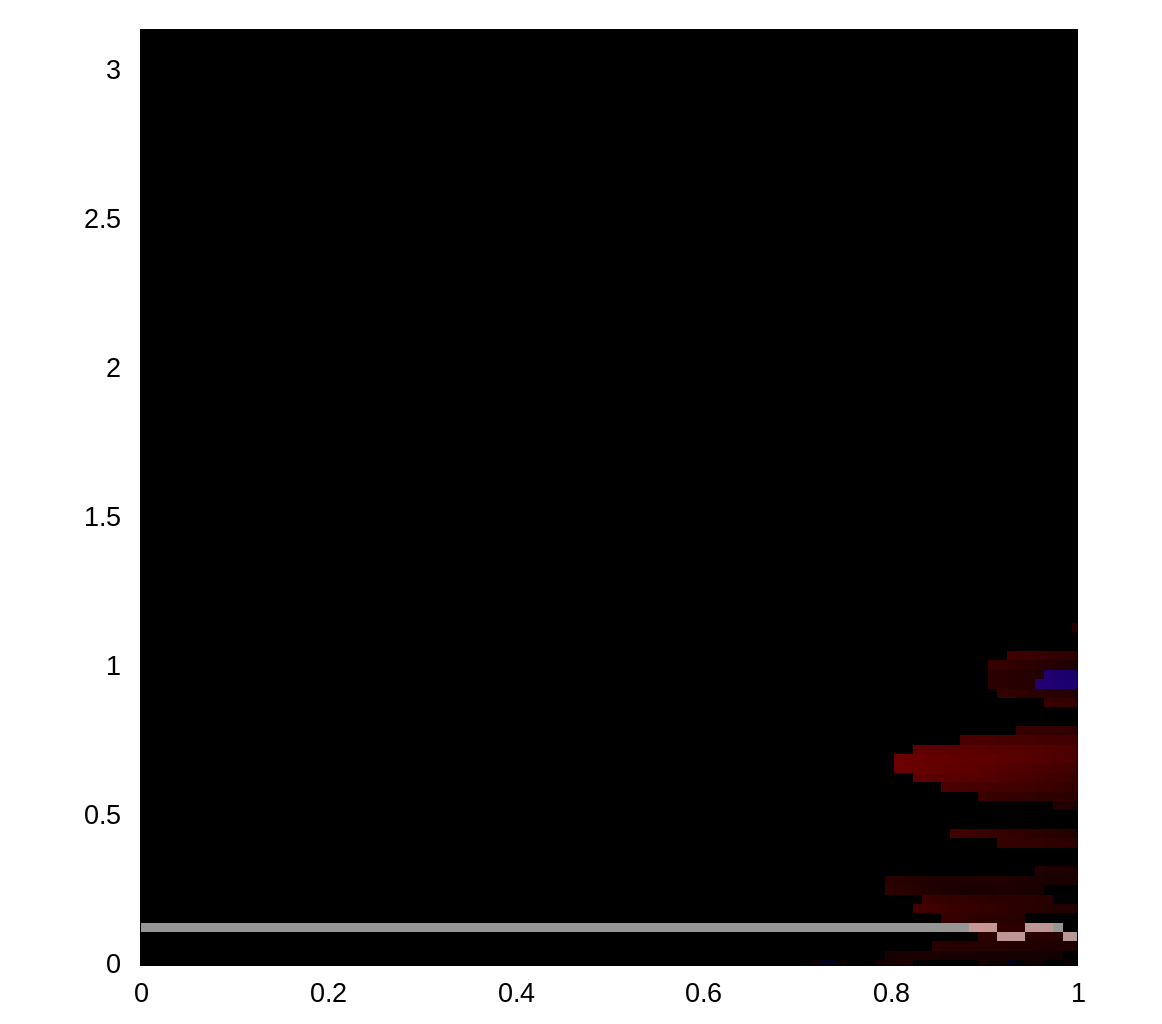} 
\includegraphics[height=1.5in]{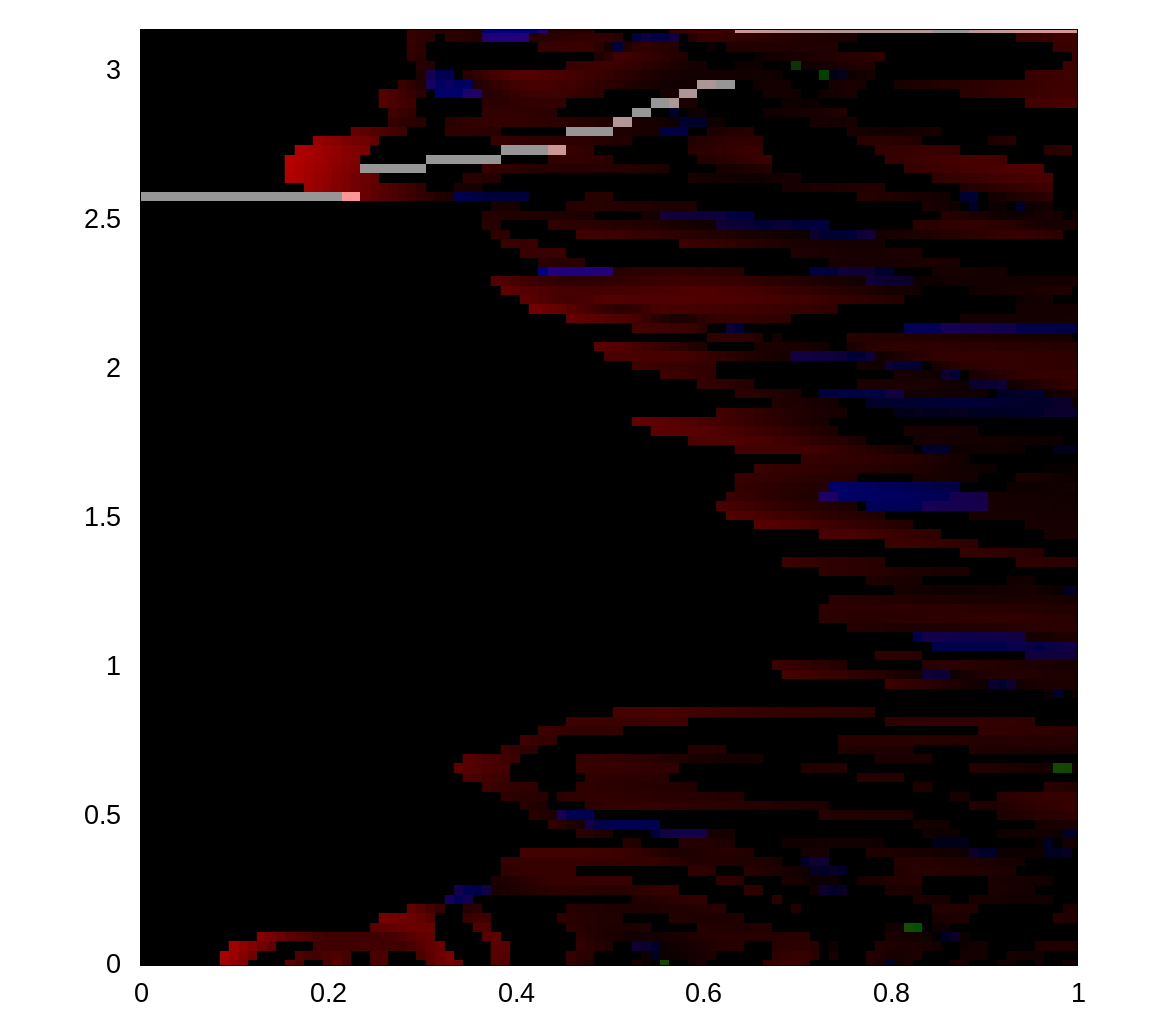} 
\includegraphics[height=1.5in]{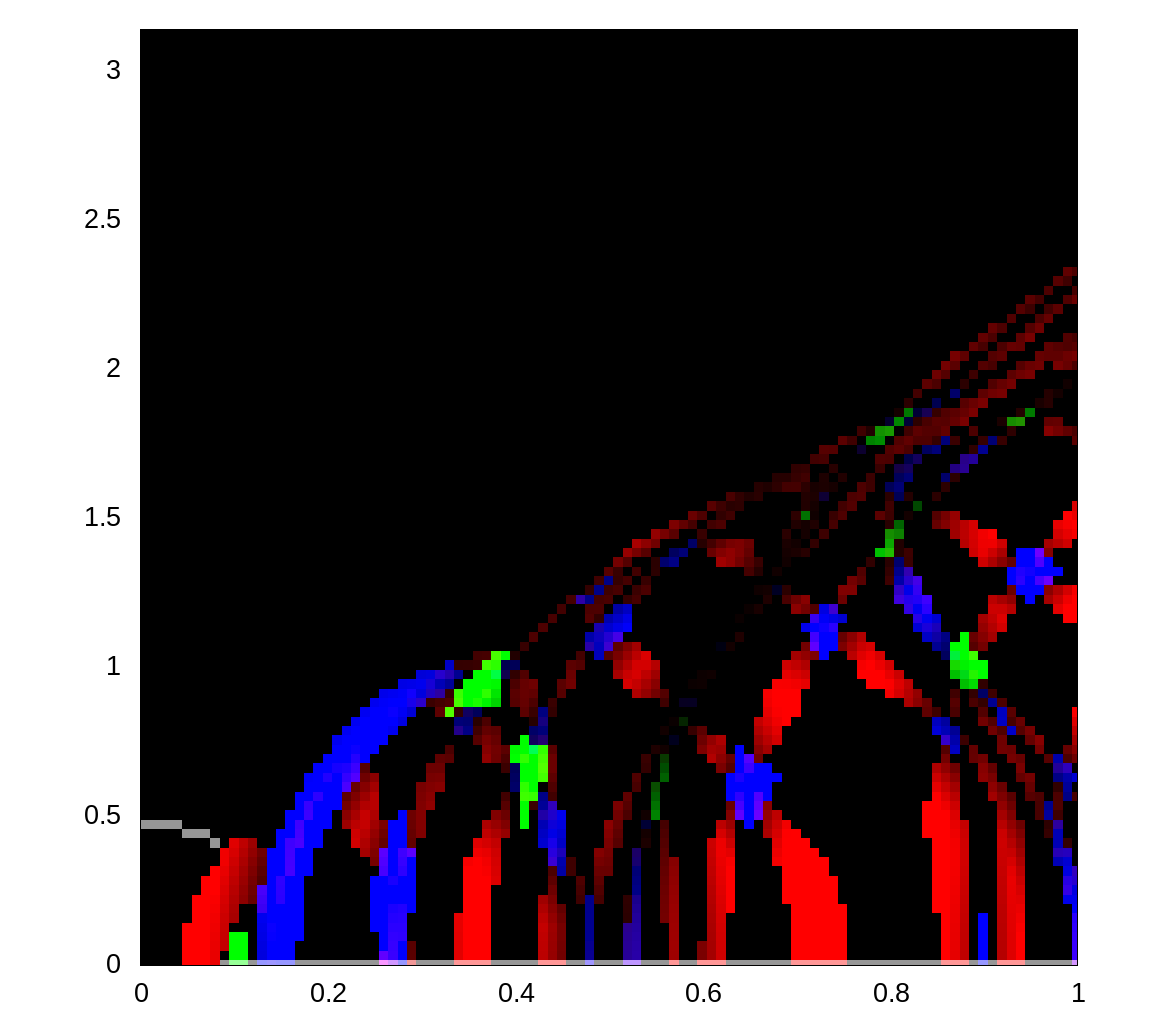} 
\includegraphics[height=1.5in]{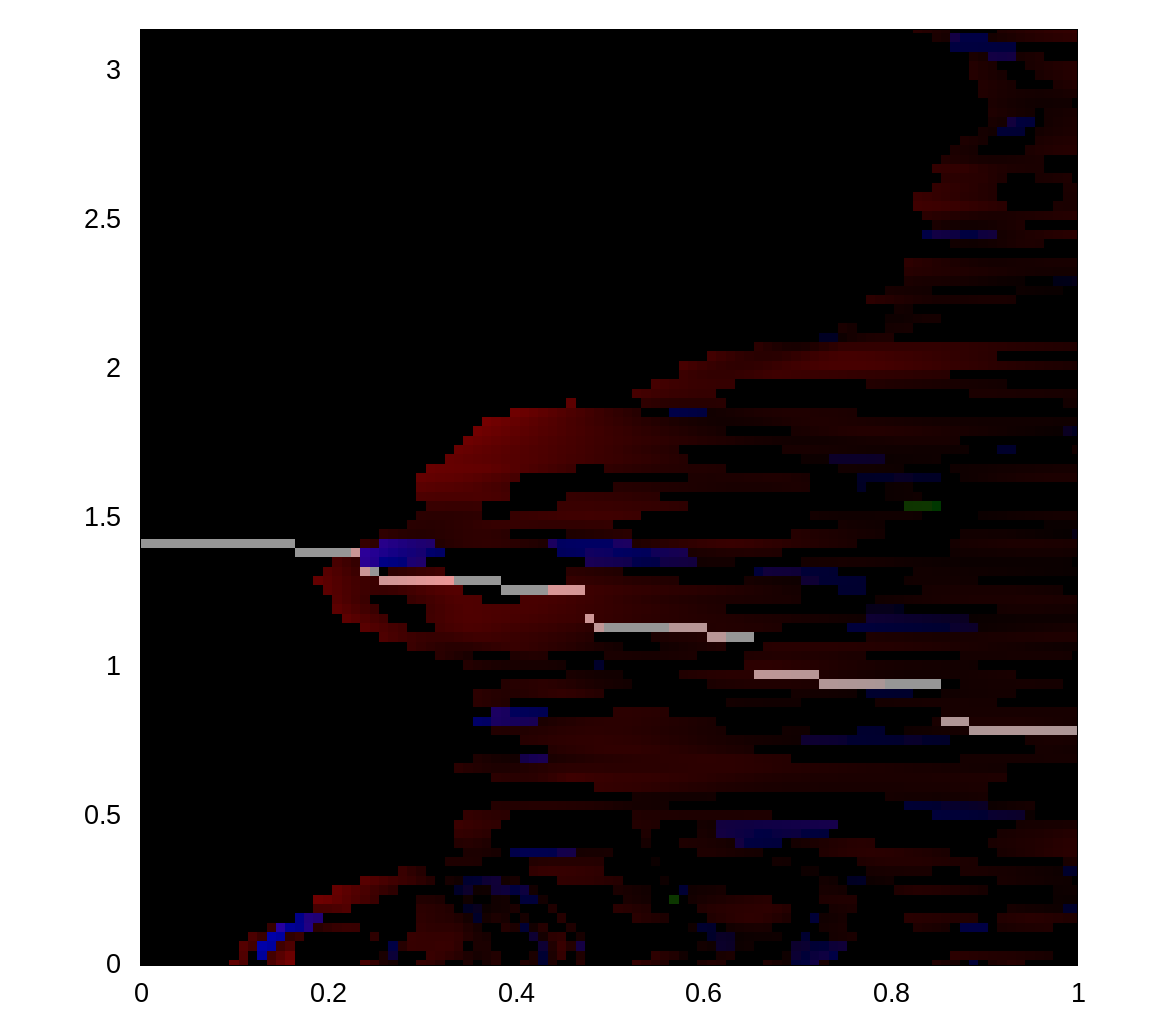} 
\hskip 0.9in (e)
\hskip 1.55in (f)
\hskip 1.55in (g)
\hskip 1.55in (h)
\caption{ 
The phase diagram of the second device.  (a,b,c,d) The isolated magnetic atoms
form a uniform line in the middle of a superconducting strip of size $60a\times
24a$.  The separation of the  magnetic atoms is $a$ (see Fig. \ref{step}a).
(e,f,g,h) The isolated magnetic atoms form a line on the edge of a
superconducting strip of size $80a\times 18a$.  The  magnetic atoms have a
random distribution, with a density $0.7$ per site (see Fig. \ref{step}b).  The
horizontal axis is the MEC $J$ and the vertical axis is the spiral wave-vector
$k^M_x a$.  The red area is the TSC phase with one MZM at an end of the wire.
The green area is the TSC phase with three MZMs.  The blue area is the SPT
phase with two MZMs.  The brightness of the color represents the gap of the
TSC, with the full brightness corresponds to the parent superconducting gap
$\Del$.  The white line marks the spiral wave-vector $k^M_x$ that corresponds
to the ground state.  The chemical potential $\mu$, the local shift of the
chemical potential $\del \mu = -V$, and the number of magnetic atoms per unit
cell $n$ are given by (a) $(\mu,\del \mu)=(-3,0)$; (b) $(\mu,\del \mu)=(-3,2)$;
(c) $(\mu,\del \mu)=(-5,2)$; (d) $(\mu,\del \mu)=(-1,2)$; (e) $(\mu,\del
\mu)=(-3,0)$; (f) $(\mu,\del \mu)=(-3,2)$; (g) $(\mu,\del \mu)=(-4.5,2)$; (h)
$(\mu,\del \mu)=(-1,2)$. 
}
\label{second} 
\end{figure*}

In the second device (see Fig. \ref{step}a),  we use a \emph{layered
superconductor} (such as FeSe) or a mon-layer superconductor.  We put a line of
\emph{isolated} magnetic atoms on the superconducing surface, where the
separation of the magnetic atoms is less than the superconducting coherence
length.  We require the magnetic atom to be ionized and cause a local shift of
chemical potential $\del \mu$ in the superconducting layer under the magnetic
atoms.  We find that if the shift of the chemical potential is of the order of the
Fermi energy, then even for a weak MEC (much less than the bandwidth), a spiral
magnetic order of the magnetic atoms will develop automatically to produce an
1D TSC with a large gap (close to the parent superconducting gap)
(see Fig. \ref{second}b,c,d,f,g,h).

In \Ref{NDL1446,PM150506078}, a device formed by a mono-atomic chain of Fe
atoms on the surface of superconducting Pb was fabricated.  The Fe atoms touch
each other and form a metallic wire as described in \Ref{KD14017048}, assuming
the Fe-chain is very long.  For this first class of devices, a strong MEC is
not required and a weak MEC is enough to induce a TSC.  However, in the
experiment, the Fe-chain is very short (less than the superconducting coherent
length) and MZMs at the chain ends, if any, may not be isolated. 
In this paper, we consider the case where the magnetic atoms do not touch each
other and form an insulating wire. We will use Yu-Shiba-Rusinov states induced
by MEC to form the 1D TSC.

In \Ref{ZK171010701}, quantized resonance conductance for a MZM was observed.
The energy scale that protects the MZM is about $T=0.1$K, which is much less
than the energy scale for the parent superconductor $T_c =1.2$K.  In this
paper, we try to design devices to produce the 1D TSC whose energy gap is close
to that of the parent superconductor (see Fig.  \ref{first}).  For a quantum
wire made by FeSe monolayer on SrTiO$_3$, such a gap can be as large as
$4$meV$\sim 40$K.

\noindent \textbf{Model for the first device:} 
Since the 1D SC wire is formed a monolayer film or a layered material, if we
ignore the interlayer hopping, we can model such a wire covered magnetic atoms
by a 2D model on a square lattice of size $L_x\times L_y$ (see Fig. \ref{first}):
\begin{align}
\label{HtDel}
 H  = &
-\sum_{\v i,\del \v i} t_{\v \mu} c_{\v i+\del \v i,\al}^\dag c_{\v i,\al} 
+\sum_{\v i} [\Del (c_{\v i \up} c_{\v i\down}-c_{\v i \down} c_{\v i\up}) +h.c.]
\nonumber \\
&
-\sum_{\v i} \mu c_{\v i}^\dagger c_{\v i}
+\sum_{\v i_\text{m}} (\v B_{\v i_\text{m}}  c_{\v i_\text{m}}^\dag \v \si c_{\v i_\text{m}} 
+ V c_{\v i_\text{m}}^\dagger c_{\v i_\text{m}})
,
\end{align}
where $\del \v i =\pm \v x,\pm \v y$, and $\sum_{\v i_\text{m}}$ sums over the
sites with the magnetic atoms.  $\v B_{\v i}=J \v n_{\v i}$ describes
the MEC to the magnetic atoms where the unit vectors $\v n_{\v i}$ describe the
spin orientation of the magnetic atoms.  $V$ describes potential shift caused
by the magnetic atoms.  We will choose $t_x=t_y=1$, $\mu=-3$ (\ie $E_F=1$), 
$\Del=0.05$, and $(L_x,L_y)=(100,10)a$ where $a$ is the lattice constant.
Such a system has a generic circular Fermi surface. 
The corresponding superconducting coherent length is $ \xi =
\frac{\hbar v_F}{\pi \Del} =10a$.
The above choice of parameters roughly models the FeSe monolayer.

We further assume the magnetic atoms to have a spiral magnetic order
\begin{align}
\label{Bsp}
\v B_{\v i}  = J \Big(\sin(k^M_x i_x a),0,\cos(k^M_x i_x a)\Big) .
\end{align}
where $k^M_x$ is spiral wave-vector.  $k^M_x$ is a dynamical parameter. The
ground state has a $k^M_x$ that minimizes the total energy.

Assuming one magnetic atom per site with $V=0$ (\ie a
uniform distribution), we find a phase diagram in Fig.  \ref{first}b via a
numerical calculation of the energy spectrum of \eqn{HtDel}.  
(The uniform case is also studied in \Ref{SB14101734}.)
The Yu-Shiba-Rusinov states \cite{Y6575,S6835,R6985} induced by the MEC of
magnetic atoms have a size of $\xi$, whose centers have a separation $\xi/10$.
So the Yu-Shiba-Rusinov states strongly overlap with each other.  Those
overlapping Yu-Shiba-Rusinov states can form a TSC if the spins of the magnetic
atoms form a proper spiral order.  In Fig.  \ref{first}b, we see that a
spiral order with $k^M_x$ marked white is spontaneously generated to minimize
the ground state energy (see 
Supplementary Materials Section I). For such a spiral order,
the Yu-Shiba-Rusinov states automatically form the TSC with MZMs protected by a
large energy gap.  Even in the limit $\xi \gg a$, we find the gap of the TSC
can be close to the gap of parent superconductor.  For example, for a weak MEC
$J=0.1$, the gap of TSC can reach $0.92 \Del$. 

The spiral order that minimizes the ground state energy usually produce a TSC.
This is because the Yu-Shiba-Rusinov states correspond to spinless electrons.
When 1D spinless electrons want to form a gapped superconducting state to
minimize the ground state energy, the $p$-wave superconductivity is the only
option.  We also like to remark that the number of ``wedges'' in the  phase
diagram for small $J$ is the number of 1D subbands, which is given by $k_F
L_y/2\pi$, where $k_F$ is the Fermi wave-vector.  Thus the SC wire should be
very narrow in order not to have too many subbands.

Our model actually has a twisted time reversal symmetry $T^*$, generated by the
usual time reversal and 180$^\circ$ spin-$S_y$ rotation (assuming the magnetic
order is in $xz$-plane). The $T^*$ time reversal symmetry is described by
$Z_2^T$ symmetry group.  The 1D fermionic topological phases with $T^*$ time
reversal symmetry are classified by $\Z_8$ integers\cite{FK1009} $n_{MZM}$ --
the number of the MZMs at one end of the wire  (see 
Supplementary materials Section II).  The odd-$n_{MZM}$ phases correspond to
iTO, and the even-$n_{MZM}$ phases correspond to symmetry protected topological
(SPT) orders.\cite{GW0931}

The TSC phase is also robust against disorder in the local density of the
magnetic atoms.  Fig. \ref{first}c is the phase diagram for $\Del = 0.05$,
$\mu=-3$, $V=0.5$, $(L_x,L_y)=(100,10)a$ and there are 0.3 magnetic atoms per
site. (The percolation threshold is 0.592 atoms per site.)  Again, for the
spiral order that minimizes the ground state energy, the Yu-Shiba-Rusinov
states automatically form the TSC with MZMs.  For a weak MEC $J=0.3$, the gap
of TSC is about $0.4 \Del$.  Fig.  \ref{first}d is the phase diagram for
even stronger potential shift $V=t_x$.

We see that the randomness in the MEC and the chemical potential
caused by the random distribution of the magnetic atom increase the required MEC
and reduce the gap of TSC.  But even with that randomness, the required MEC is
still quite weak: $J=0.25$ (for $V=0.5$).

The system in Fig. \ref{first} models a SC wire whose width is 10 lattice
spacing (a few nanometers).  The SC coherent length $\xi$ is also a few
nanometers. (But a longer SC coherent length will not significantly affect our
results, see Supplementary Materials Fig.  \ref{thBkO}:Right.) The wire is covered by magnetic atoms
with a density of $0.3$ per site, which is well below the percolation threshold of
0.592 atoms per site.  In fact, the above parameters fit a SC wire made by FeSe
superconducting thin film quite well.

In the following, we summarize some key factors in the device design to
realize the 1D TSC with MZMs:
\begin{enumerate}
\item 
A SC wire is formed by a monolayer thin film, or quasi-2D material 
(such as FeSe superconductor with $\xi=5.1$nm$=14a$), or material
whose Fermi surface contains tube-like parts or parallel sheets.
\item
The parent superconductor is an $s$-wave superconductor with no spin-orbital
coupling, or the spin-orbital coupling still conserves one component of the
spin.
\item
The width of the SC wire is a few times $2\pi/k_F$, which is less than the SC
coherent length $\xi$.  
\Ref{LC10060420} has developed a method to make nano-wires as narrow as $5$nm.
\item
The SC wire is covered by isolated magnetic atoms that do not touch each other.
Note that the distribution of the magnetic atoms can be random, as long as
their density is much higher than $1/\xi^2$.  
\item
The magnetic atoms have a weak spin-orbital coupling
less than the superconducting gap $\Del$, so that the spins can rotate freely.
\item
If the magnetic atoms have a random distribution, they should not cause a too
large local chemical potential shift $\del \mu=-V$.  If the magnetic atoms have
a uniform distribution, then a large local chemical potential shift $\del
\mu=-V$ may help to produce the 1D TSC for weak MECs.  In this case, we do not
need a SC wire, since the magnetic atoms can make an effective SC wire by
themselves.
\end{enumerate}
The resulting device can realize TSC with a sizable gap ($\sim 0.4$ of parent
superconducting gap), even for a weak MEC ($J \sim 0.25$).

\noindent \textbf{Model for the second device:} 
However, it is hard to make a very narrow superconducting wire.  To solve this
problem, we propose a second device (see Fig. \ref{step}a).  In this device, we
place a line of \emph{isolated} magnetic atoms on the surface of a layered
superconductor, such as FeSe.  But now we require the magnetic atoms to cause a
large local chemical potential shift $\del \mu=-V$.  In this case, the magnetic
atoms will induce an effective SC wire under them, which may lead to a 1D TSC
state.  We still use \eqn{HtDel} to model our device.  We will assume
$(L_x,L_y)=(60,24)$, $t_x=t_y=1$, and $\Del=0.05$.  We still assume the
magnetic atoms to have a spiral magnetic order described by \eqn{Bsp}.  The
resulting phase diagram is given in Fig.  \ref{second}a,b,c,d.  (Note that, on
the FeSe surface, magnetic atoms separated by one lattice spacing can still be
isolated.)

From Fig. \ref{second}a, we find that if the magnetic atoms do not dope the
superconductor (\ie the local shift of the chemical potential $\del \mu=-V=0$),
then the TSC phase cannot be realized by the ground state of the spin spiral
state, for weak MEC $J < 1$.  From Fig.  \ref{second}b,c,d, we find that if the
magnetic atoms dope the superconductor (\ie the local shift of the chemical
potential $\del \mu=-V \sim t_x$), then the ground state of the spin spiral
state also realize the TSC phase for weak MEC $J \sim 0.1$.  The induced 1D TSC
state has a gap close to that of the parent SC state.  We like to remark that
if the FeSe layers have steps on the surface, placing the magnetic atoms along
a straight step (see Fig. \ref{step}b) will help to produce the 1D TSC state.

In  Fig. \ref{second}b, the local doping enlarge the electron-like Fermi
surface.  In  Fig. \ref{second}c, the local doping creates an electron-like
Fermi surface.  In  Fig. \ref{second}d, the local doping change the
electron-like Fermi surface to a hole-like Fermi surface.  In those three
cases, we see that the local doping from the magnetic atoms can produce an
effective SC wire on a surface of layered SC.  The spin of the magnetic atoms
will develop a spiral order to realize the TSC phase even for a weak MEC (about
1/5 of the Fermi energy 1/20 of the bandwidth).  The gap of the STC is about
that same as the parent SC gap.

In Fig. \ref{second}e,f,g,h, we also consider a device where the magnetic
atoms form a line at a step edge of the SC thin film.  Here we allow the
electron to have a random distribution with each occupied by 1 or 0 atoms.  The
average density is $0.7$ atoms per site.  In  Fig. \ref{second}f, the local
doping enlarge the electron-like Fermi surface.  In  Fig. \ref{second}g, the
local doping creates an electron-like Fermi surface.  In  Fig. \ref{second}h,
the local doping change the electron-like Fermi surface to a hole-like Fermi
surface.  In those three cases, the magnetic atoms can still realize the TSC
phase for a weak MEC (about 1/5 of the Fermi energy 1/20 of the bandwidth).
The gap of the STC can be as large as 1/2 of the parent SC gap.  
We would like
to remark that if the SC thin film, such as FeSe layers have steps on the
surface, placing the magnetic atoms along a straight step (see Fig.
\ref{step}b) will help to produce the 1D TSC state, which will be even stronger
than that shown in Fig. \ref{second}b,c,d.

We would like to thank A. Chang, J.-F. Jia, P.A. Lee, J.-W. Mei, J.S. Moodera,
G. Wang, H.-H. Wen, and Q.-K. Xue for helpful discussions.  This work is
supported by NSF grant DMR-1506475 and DMS-1664412.

\vfill
\eject


\centerline{\large \bf
Supplementary Materials
}

\section{Ground state energy of a superconductor} \label{ground}

In general, a superconductor is described by  
\begin{align}
H & = \sum_{I J}  c_I^\dagger t_{IJ}c_J+\sum_{I<J} \left(c_I\Delta_{IJ}c_J+h.c. \right)
\nonumber\\
& = \sum_{I J}  \left[c_I^\dagger t_{IJ}c_J+ \left(c_I\frac {\Delta_{IJ}}{2} c_J+h.c. \right) \right],
\end{align}
where $I$ labels the electron spin and site position.
In Majorana basis
\begin{align}
\begin{split}
&\begin{pmatrix}
c_I \\
c_I^\dagger
\end{pmatrix}=\frac{1}{\sqrt{2}}\begin{pmatrix}
1 & \ii \\
1 & -\ii 
\end{pmatrix}\otimes \Id_{2N_\text{site}}\begin{pmatrix}
\gamma_{I,1}\\
\gamma_{I,2}
\end{pmatrix}\\
&\{\ga_{I,a},\ga_{J,b}\}=\del_{IJ}\del_{ab},
\end{split}
\end{align}
$H$ becomes
\begin{align}
H
= &\sum_{I J}\left[ \frac12 (\gamma_{I,1}-\ii \gamma_{I,2}) t_{IJ}(\gamma_{J,1}+\ii \gamma_{J,2})\right.
\nonumber\\
&\ \ \ \ \ \
+\frac12 (\gamma_{I,1}+\ii \gamma_{I,2})\frac{\Delta_{IJ}}{2} (\gamma_{J,1}+\ii \gamma_{J,2}) 
\nonumber\\
&\ \ \ \ \ \
\left.+\frac12 (\gamma_{J,1}-\ii \gamma_{J,2})\frac{\Delta_{IJ}^*}{2} (\gamma_{I,1}-\ii \gamma_{I,2}) \right]
\nonumber\\
&=
\frac12  \Tr(t)
+\frac12 \sum_{IJ} (\gamma_{I,1},\gamma_{I,2})(M_{\gamma})_{IJ}\begin{pmatrix}
\gamma_{J,1}\\ \gamma_{J,2}
\end{pmatrix}
\end{align}
where
\begin{align}
M_{\gamma} 
&= \ii
\begin{pmatrix}
 \Im( t) + \Im( \Del), & \Re( t)-\Re( \Delta) \\
-\Re( t)-\Re( \Delta), &  \Im( t) - \Im( \Del)
\end{pmatrix}
\end{align}
Note that $M_\ga$ satisfies
\begin{align}
 M_\ga^\dag =M_\ga=-M_\ga^\top.
\end{align}
We can use an orthogonal matrix $O$ to transforms $H_{\gamma}$
into the canonical form 
\begin{align}
 O M_{\gamma} O^\top
=(E_a\del_{ab}) \otimes \ii \si^2 , \ \ \ E_a>0.
\end{align}
Now $H$ becomes
\begin{align}
 H &=\frac 12 \Tr(t)+
\sum_a 
E_a 
\ii \la_{a,1} \la_{a,2} ,
\nonumber \\
\la_{a,1}^2 &= \la_{a,2}^2=\frac12 ,
\end{align}
Note that
$\ii \la_{a,1} \la_{a,2}$ has eigenvalues $\pm 1/2$ and
$\ii \la_{a,1} \la_{a,2}$ with different $a$'s commute with each other.
The ground state satisfies
\begin{align}
 \ii \la_{a,1} \la_{a,2} |\text{ground}\> = -\frac12 |\text{ground}\>.
\end{align}
The groud state energy is given by
\begin{align}
E_\text{ground}=
 \frac12 [\Tr(t) -\sum_a E_a].
\end{align}

\section{The fermionic SPT phases protected by $Z_2^T$ time reversal symmetry}
\label{TSPT}

In the presence of $Z_2^T$ time-reversal symmetry,
the hopping coefficient $\v t_{\v \mu}$ and $\t \Del$ are real, and we have
\begin{align}
 \v t(-k_x)=&\v t^*(k_x)=\v t^\top(k_x) .
\end{align}
In this case
\begin{align}
& M_{\gamma}(k_x)= 
\begin{pmatrix}
0 & \v t(k_x)- \v \Delta \\
-\v t(k_x)- \v \Delta & 0
\end{pmatrix}
\end{align}
In the gapped phase $\det(\v t(k_x)- \v \Delta) \neq 0$.  Thus wind number $W$,
the number of times $\det(\v t(k_x)- \v \Delta)$ wind around $0$ as we change
$k_x=0$ to $k_x=2\pi$ is well defined.  Such a winding number $W$ characterizes
the $Z_2^T$-SPT order of the gapped 1D SMI-SC junction.  The  $Z_2^T$-SPT state
will have $W$ number of MZM's at an end of the 1D SMI-SC junction.

\section{Some other device designs to produce MZMs}
\label{other}

In this section, we consider some other device designs that can produce MZMs.
We will consider junction between a stripe of spiral magnetic insulator (SMI)
and an $s$-wave superconductor.  We find that, in general, that those devices
can give rise to the 1D TSC phase only when the magnetic exchanging coupling is
of order the bandwidth of the SC. 
In the following, we will concentrate on how to design devices that can realize
the 1D TSC phase with a gap close to the SC gap at small magnetic exchanging
couplings.

\begin{figure}[tb] 
\centering \includegraphics[scale=0.6]{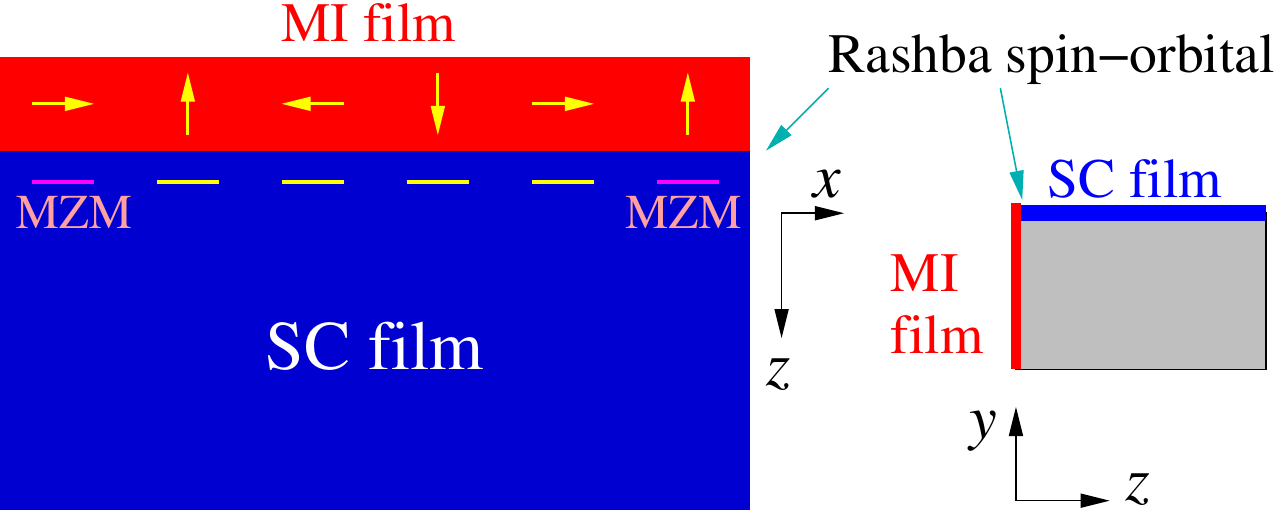} 
\caption{ A junction between a magnetic insulator thin film and an $s$-wave superconductor
thin film.  The pair-breaking MEC from the magnetic insulator can induce energy
levels inside the SC gap.  The hopping of the Majorana fermions (the Bogoliubov
quasiparticles) between those levels can realize a 1D TSC with topological
degeneracies (\ie MZMs for non-interacting cases) at the chain ends.  If the
magnetic insulator is a SMI, TSC can appear without Rashba spin-orbital
coupling.  If the magnetic insulator is a ferromagnetic insulator, the appearance of TSC requires
the presence of a Rashba spin-orbital coupling on the SC surface.  
}
\label{SCMI} 
\end{figure}

\begin{figure}[tb] 
\centering \includegraphics[scale=0.6]{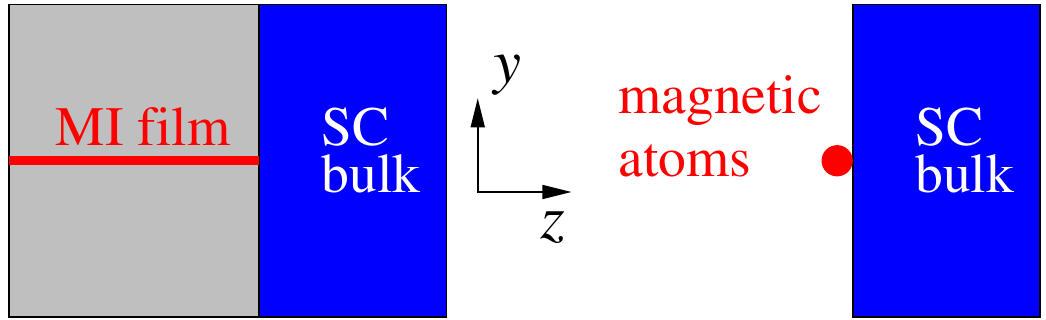} 
\caption{ 
A junction between a SC bulk sample and a magnetic insulator stripe on
the SC surface.  The SC surface is the $x$-$y$ plane and the magnetic stripe is
in $x$-direction.  The  magnetic insulator stripe may also be replace by an 1D
array of magnetic atoms that do not touch each other.
}
\label{SCbMI} 
\end{figure}

\begin{figure*}[tb] 
\centering 
\includegraphics[height=1.5in]{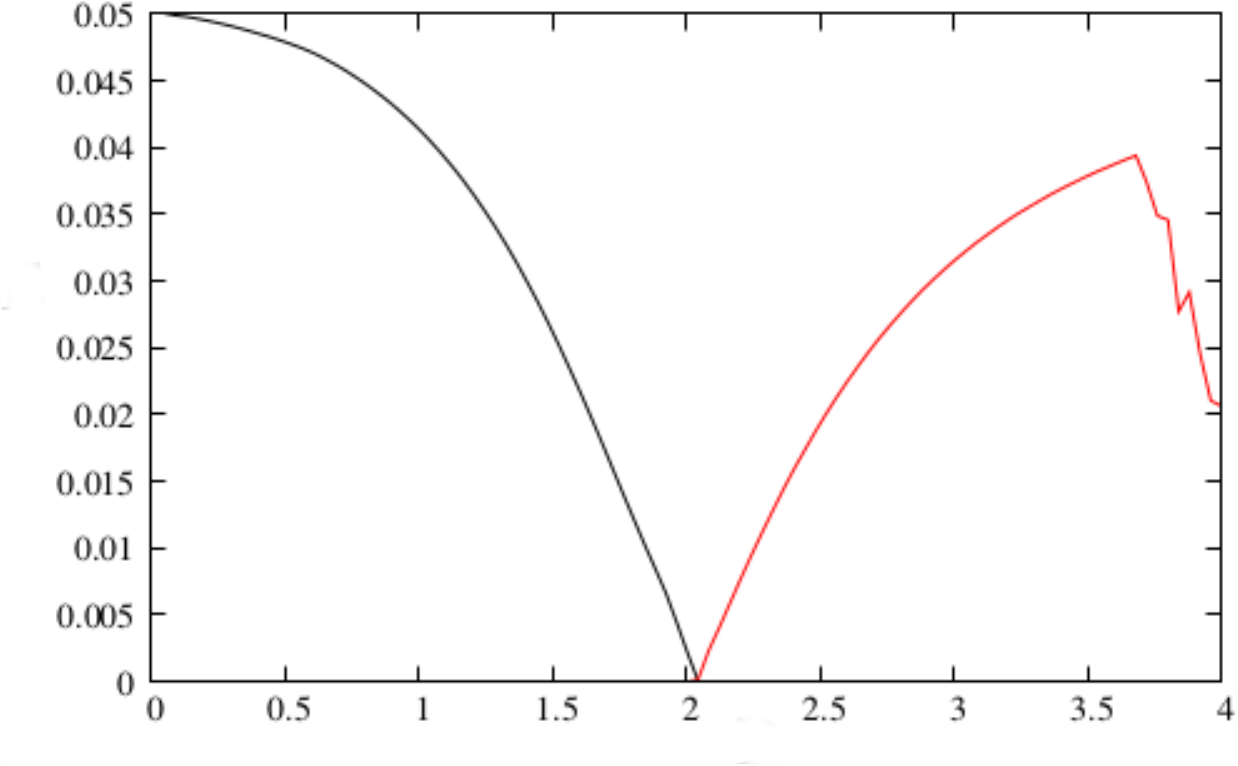}
~~~~~~~~~~~
\includegraphics[height=1.5in]{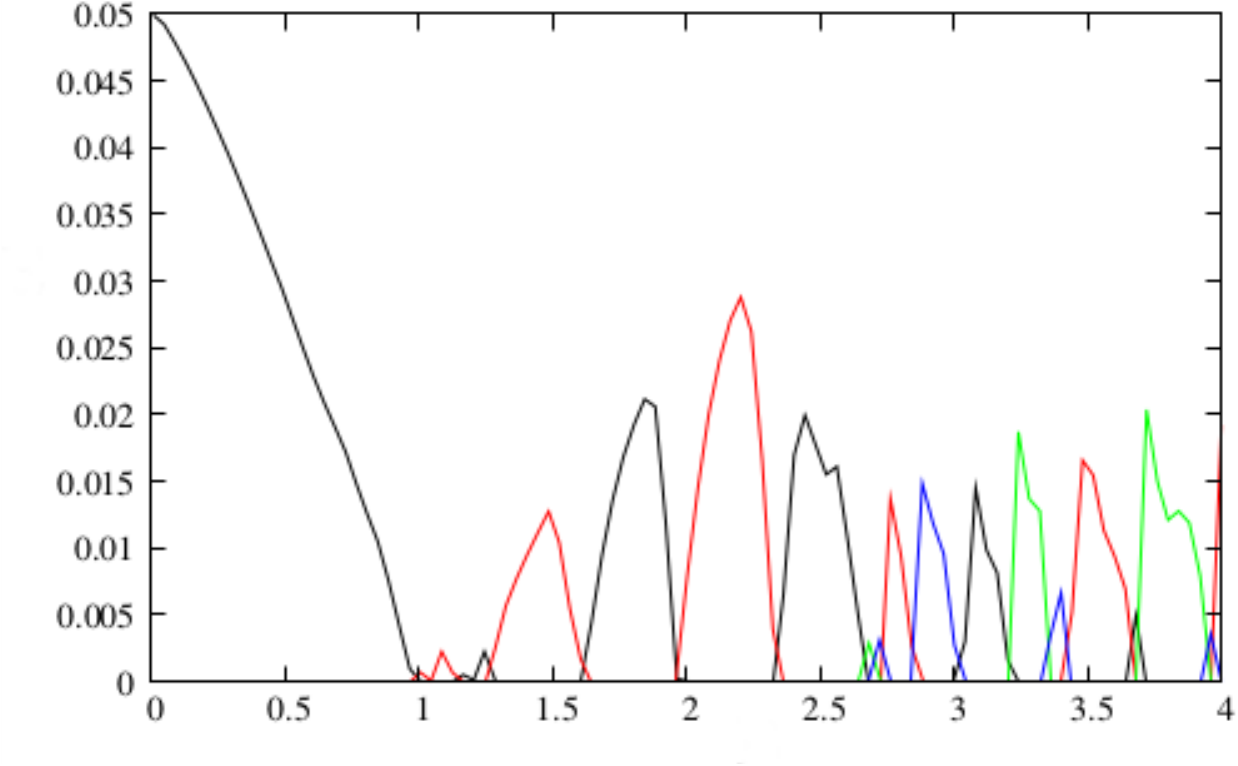}
\caption{
The lowest positive one-particle energy level $E_1$ of $H$ in \eqref{HtDel3D}
with hopping $t_x=t_y=t_z=1$, SC gap $\Del=0.05$, and chemical potential
$\mu=-4$, for a bulk SC coupled to a SMI stripe in Fig. \ref{SCbMI}.  The spiral
wave vector of SMI is $\v k^M=\frac{2\pi}{3a} \v x$.  The horizontal axis is
the MEC $J$. The width of the SMI stripe is  Left:  $d_y=a$
and Right: $d_y=20a$.  The red and green curves represent the TSC phase (with
$\text{Pf}[M_\ga(0)]<0, \text{Pf}[M_\ga(\pi)]>0$ for red and
$\text{Pf}[M_\ga(0)]>0, \text{Pf}[M_\ga(\pi)]<0$ for green).  The black curves
represent the trivial gapped phase (with $\text{Pf}[M_\ga(0)]>0,
\text{Pf}[M_\ga(\pi)]>0$).  The blue curves represents the SPT phase protected
by translation symmetry (with $\text{Pf}[M_\ga(0)]<0, \text{Pf}[M_\ga(\pi)]<0$).
}
\label{SCbSMI} 
\end{figure*}

\subsection{Two coupled ferromagnetic dots}

First, we note that a magnetic atom at the surface of superconductor induces an
energy level in the superconductor below the SC gap.  The energy level has its
spin pointing in $\v n_{\v i}$ direction.  Two magnetic atoms will induce two
energy levels inside the SC gap.  Two nearby energy levels with spins $\v n_i$
and $\v n_j$ have the following form of coupling which contains both hopping
$t_{\v n_i,\v n_j}$ and pairing $\Del_{\v n_i,\v n_j}$ terms:
\begin{align}
H &= 
\eps_0 (\psi^\dag_{\v n_i} \psi_{\v n_i}+\psi^\dag_{\v n_j} \psi_{\v n_j})
\nonumber\\
&
+
t_{\v n_i,\v n_j}  \psi^\dag_{\v n_i} \psi_{\v n_j}
+
\Del_{\v n_i,\v n_j}  \psi_{\v n_i} \psi_{\v n_j}
+h.c.
\end{align}
The spin-$\v n$ wave function is given by
\begin{align}
 \vphi_{\v n} = 
\bpm
\ee^{\ii \frac \phi 2} \cos \frac \th 2 \\
\ee^{-\ii \frac \phi 2} \sin \frac \th 2 \\
\epm
\end{align}
where $\v n = (\sin \th \cos \phi, \sin \th \sin \phi, \cos \th)$.
We find
\begin{align}
&\ \ \ \ t_{\v n_i,\v n_j} = t \vphi_{\v n_i}^\dag \vphi_{\v n_j}
\nonumber\\
&= t(
\ee^{\ii \frac{\phi_i-\phi_j}{2}} \cos \frac {\th_i}{2}\cos \frac{\th_j}{2}
+\ee^{-\ii \frac{\phi_i-\phi_j}{2}} \sin \frac {\th_i}{2}\sin \frac{\th_j}{2})
\nonumber\\
&\ \ \ \ \Del_{\v n_i,\v n_j}  = \Del \vphi_{\v n_i}^\top \ii \si^y \vphi_{\v n_j}
\\
&= \Del(
\ee^{\ii \frac{\phi_i-\phi_j}{2}} \cos \frac {\th_i}{2}\sin \frac{\th_j}{2}
-\ee^{-\ii \frac{\phi_i-\phi_j}{2}} \sin \frac {\th_i}{2}\cos \frac{\th_j}{2} )
\nonumber 
\end{align}
When $\phi_i=0$, the above becomes
\begin{align}
 t_{\v n_i,\v n_j} &= t \cos \frac {\th_i-\th_j}{2}
\nonumber\\
\Del_{\v n_i,\v n_j} \ &= -\Del \sin \frac {\th_i-\th_j}{2} .
\end{align}

We see that the strength of the hopping coupling $t_{\v n_i,\v n_j}$ and the
$p$-wave pairing coupling $\Del_{\v n_i,\v n_j}$ can be tuned by the angle
between the two spins on the two dots.  A uniform hopping coupling and $p$-wave
pairing coupling can be obtained with spiral magnetic order. This motivates us
to consider a SC thin film coupled to a spiral magnetic order.

\subsection{A SC sample coupled to a spiral  magnetic insulator}

Let us first consider the device in Fig. \ref{SCMI}, which is described by the
model \eqn{HtDel3D}.  We introduce the following site-dependent spin-$S_y$
rotation
\begin{align}
 \t c_{\v i} = \ee^{-\ii \v k^M \cdot \v i \frac{\si^y}{2}} c_{\v i}
\end{align}
to change the Hamiltonian to
\begin{align}
\label{Htc}
 H &=\sum_{\v i,\v \mu} 
[ \t c_{\v i+\v\mu}^\dag \v t_{\v \mu} \t c_{\v i} 
+ \t c_{\v i +\v\mu} \v \Delta_{\v \mu} \t c_{\v i} +h.c.]
\end{align}
where
\begin{align}
\v t_{\v \mu} =&
-t_{\v \mu}\cos \left(\frac{\v k^M\cdot \v \mu}{2}\right) \Id_2-t_{\v \mu}\sin \left(\frac{\v k^M\cdot \v \mu}{2}\right)  \ii \sigma^y
\nonumber \\
&+\delta_{\v \mu,\v 0} (\mu\Id_2+B_{i_z}\sigma^x)
\nonumber \\
\v \Delta_{\v \mu}=&\delta_{\v \mu,\v 0}\Delta_{SC} \ii \sigma^y
\end{align}
The model has a translation symmetry in $x$-direction,  
and $\v t_{\mu}$ satisfies 
\begin{align}
\v t_{-\v \mu}^\dag= \v t_{\v \mu} .
\end{align}

\begin{figure*}[tb] 
\centerline{
\includegraphics[height=2.2in]{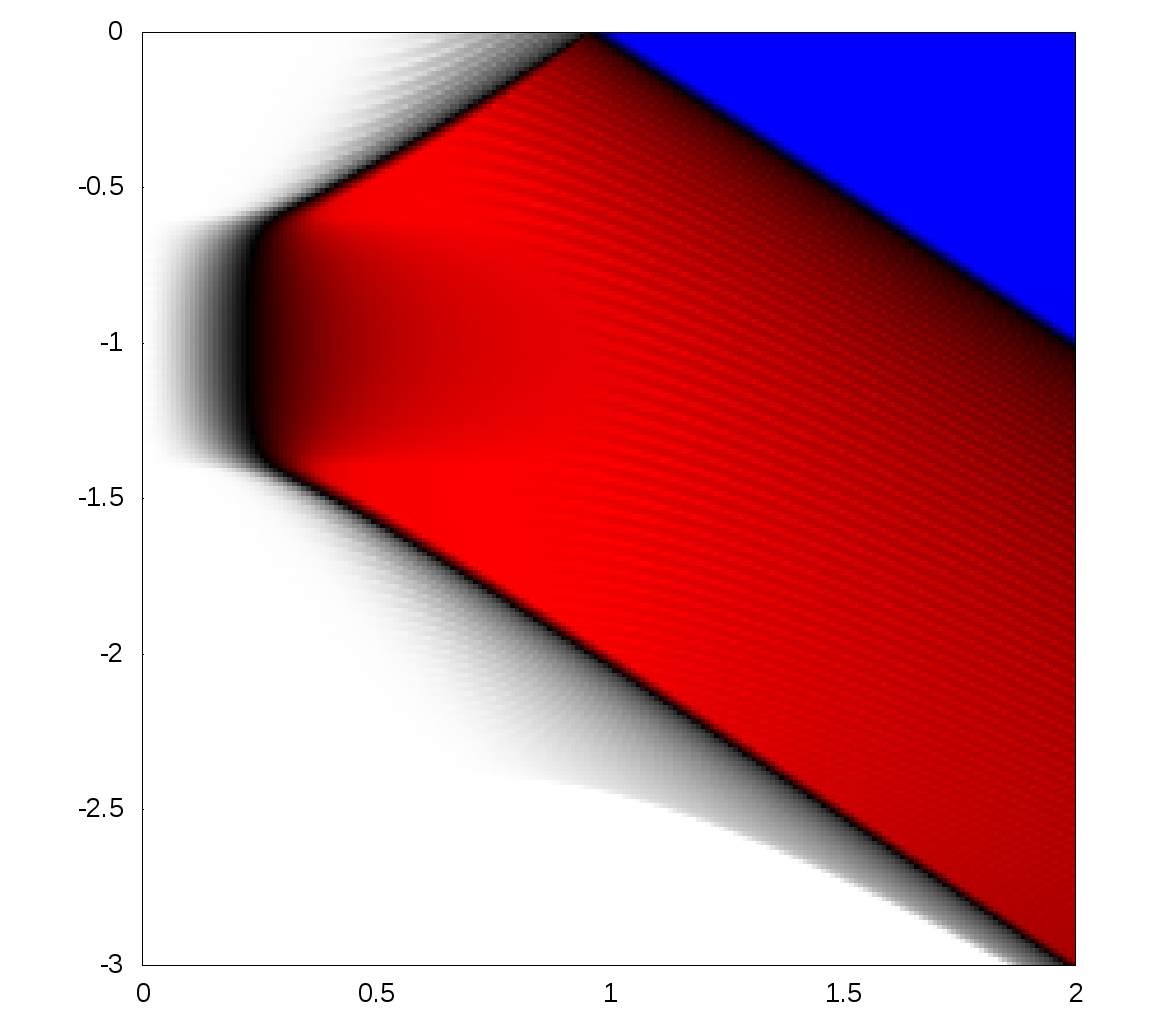} 
\includegraphics[height=2.2in]{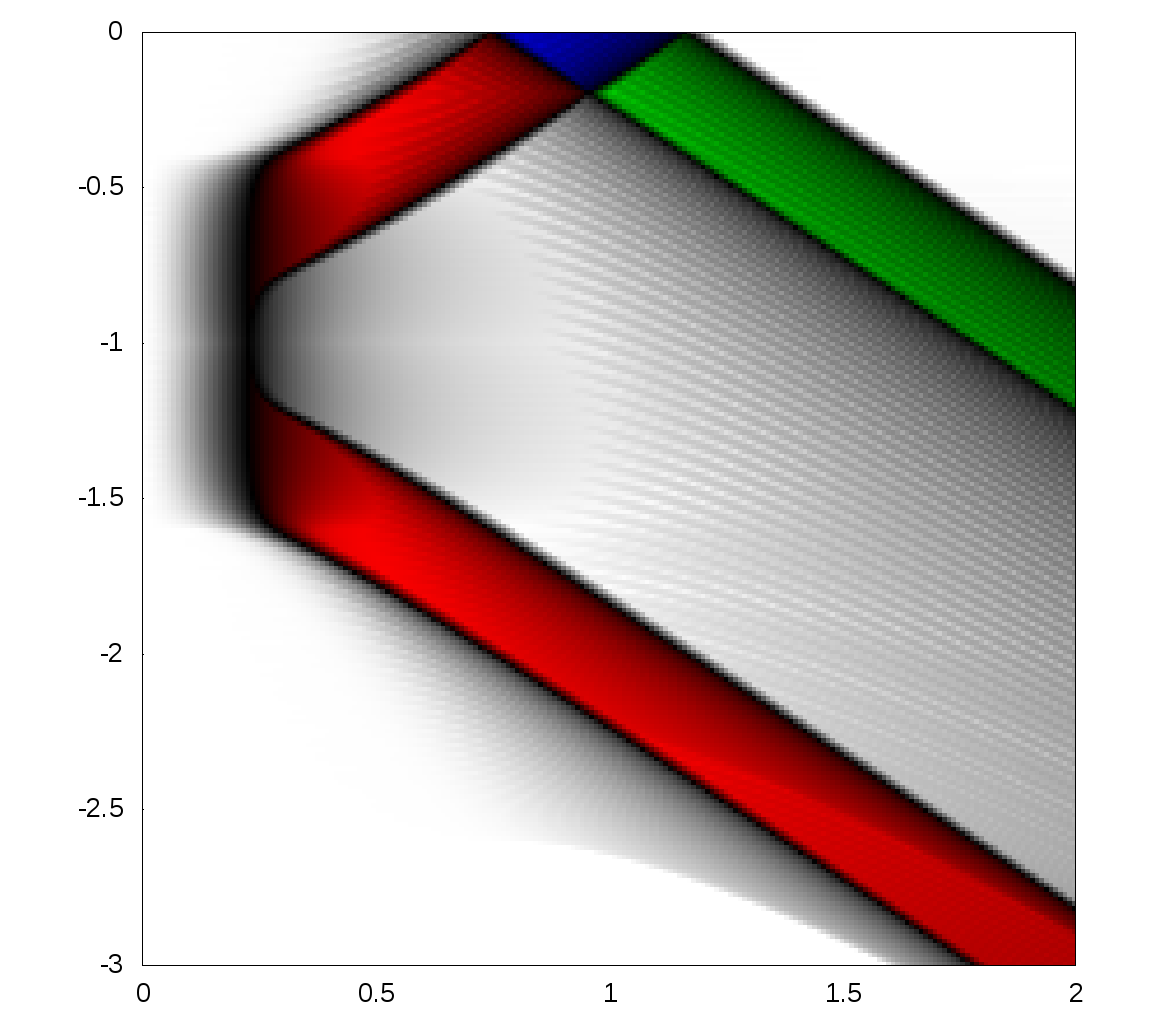} 
\includegraphics[height=2.2in]{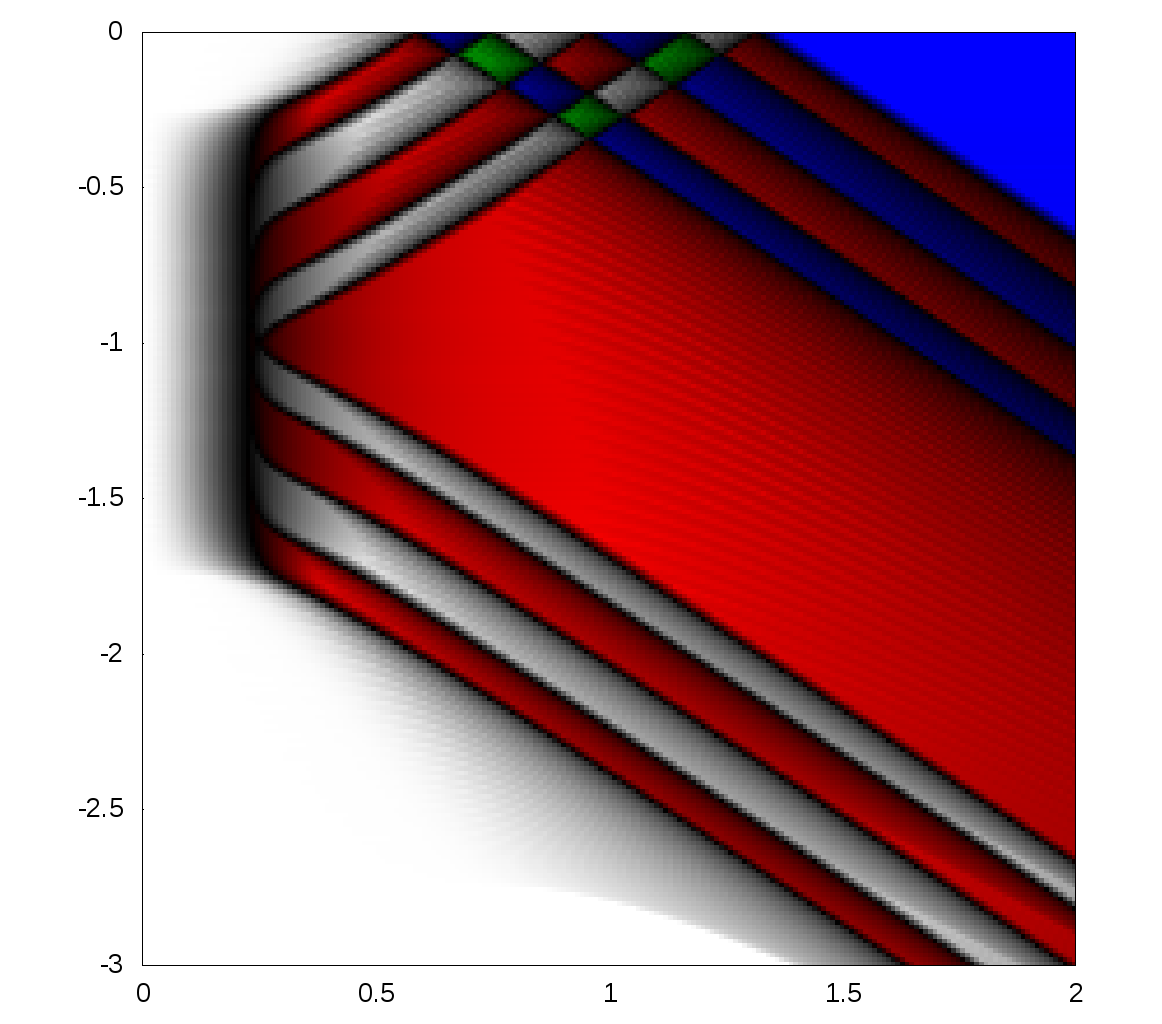} 
}
\caption{
The density plot of the lowest positive one-particle energy level $E_1$ for the
SC thin film  device in Fig. \ref{SCMI}.  The device is described by in
\eqref{HtDel3D}  with hopping $t_x=1$, $t_y=t_z=0.2$ (\ie a quasi 1D superconductor) and spiral wave vector $\v
k^M=(\frac{2\pi}{3a},0)$.   The horizontal axis is the MEC $J$ and the vertical axis
is the chemical potential $\mu$.  The brightness is proportional to $E_1$ with
maximum brightness corresponds to $E_1=\Del=0.05$.  In the above and the
similar phase diagram, the red and green areas represents the 1D TSC phase
(with $\text{Pf}[M_\ga(0)]<0, \text{Pf}[M_\ga(\pi)]>0$ for red and
$\text{Pf}[M_\ga(0)]>0, \text{Pf}[M_\ga(\pi)]<0$ for green).  The grey and the
white areas represents the trivial gapped phase (with $\text{Pf}[M_\ga(0)]>0,
\text{Pf}[M_\ga(\pi)]>0$).  The blue areas represents the SPT phase protected
by translation symmetry (with $\text{Pf}[M_\ga(0)]<0, \text{Pf}[M_\ga(\pi)]<0$).
The thickness of the SC thin film are Left: $d_y=a$; Middle: $d_y=2a$; Right:
$d_y=5a$.
}
\label{Bmu1D1on1x40a} 
\end{figure*}

After the Fourier transformation in $x$-direction, we obtain
\begin{align}
\label{Hkx}
 H  &=
\sum_{k_x,i_y,i_z} \left\{
[ \Del (\t c_{k_x,i_y,i_z}\ii\si^y \t c_{-k_x,i_y,i_z} +h.c.]\right.
\nonumber\\
&
+  
[B_{i_z}  \t c_{k_x,i_y,i_z}^\dag \si^x \t c_{k_x,i_y,i_z} +\mu \t c_{k_x,i_y,i_z}^\dagger \t c_{k_x,i_y,i_z} ]
\nonumber \\
&+  
[-t_x \t c_{k_x,i_y,i_z}^\dag 
\ee^{\ii k^M_x \frac{\si^y}{2}} 
\ee^{-\ii k_x } 
\t c_{k_x,i_y,i_z} +h.c.]
\nonumber \\
&+ 
[-t_y \t c_{k_x,i_y+1,i_z}^\dag 
\ee^{\ii k^M_y \frac{\si^y}{2}} 
\t c_{k_x,i_y,i_z} +h.c.]
\nonumber \\
&\left.+
[-t_z \t c_{k_x,i_y,i_z+1}^\dag 
\t c_{k_x,i_y,i_z} +h.c.]\right\}
\end{align}
The band structure is determined by the matrix
\begin{align}
M(k_x) &=
\bpm
\v t(k_x) & \v \Del\\
-\v \Del^* &  -\v t^*(-k_x)\\
\epm
\end{align}
where the matrices $\v t(k_x)=\v t^\dag(k_x)$ and $\v \Del=-\v \Del^\top$ can
be obtained from \eqref{Hkx}. 

In Majorana basis
\begin{align}
\begin{pmatrix}
c_I \\
c_I^\dagger
\end{pmatrix}=\frac{1}{\sqrt{2}}\begin{pmatrix}
1 & \ii \\
1 & -\ii 
\end{pmatrix}\otimes \Id_{2d_xd_yd_z}\begin{pmatrix}
\gamma_{I,1}\\
\gamma_{I,2}
\end{pmatrix}
\end{align}
$M(k_x)$ becomes
\begin{align}
&\ii M_{\gamma}(k_x)= 
U^\dag M(k_x)U 
\\
&= \ii
\begin{pmatrix}
\frac{\v t(k_x)-\v t^*(-k_x)}{2\ii} + \Im \v \Del & \frac{\v t(k_x)+\v t^*(-k_x)}{2}-\Re \v \Delta \\
-\frac{\v t(k_x)+\v t^*(-k_x)}{2}-\Re \v \Delta & \frac{\v t(k_x)-\v t^*(-k_x)}{2\ii} - \Im \v \Del
\end{pmatrix}
\nonumber 
\end{align}
where
$ U = \frac{1}{\sqrt 2} \bpm
1 & \ii \\
1 & -\ii \\
\epm$.
The energy bands are given by the eigenvalues $\pm E_n(k_x)$ of $\ii
M_\ga(k_x)$.  If the lowest band $E_1(k_x)$ is non zero for all $k_x$, then the
1D SMI-SC junction is gapped.  Such a gapped state may be in a non-trivial TSC
phase (with odd number of MZM at an end
of the 1D SMI-SC junction), or in a trivial phase without MZM.  We
can detect the TSC phase by computing Pfaffian of $M_{\ga}(k_x)$ at $k_x=0,\pi$
where $M_{\ga}(k_x)$ are real anti-symmetric matrices.  If $
\text{Pf}[M_{\ga}(0)] \text{Pf}[M_{\ga}(0)]<0 $, the gapped phase is the
TSC phase. And if $ \text{Pf}[M_{\ga}(0)] \text{Pf}[M_{\ga}(0)]>0 $, the
gapped phase is the trivial phase.

In fact in the presence of translation symmetry in $x$-direction, there are
actually four different phases, charactered by the signs of $
\text{Pf}[M_{\ga}(0)]$ and $\text{Pf}[M_{\ga}(0)]$.  The phase with $
\text{Pf}[M_{\ga}(0)]>0$ and $\text{Pf}[M_{\ga}(\pi)]>0$ is trivial, while the
phase with $ \text{Pf}[M_{\ga}(0)]<0$ and $\text{Pf}[M_{\ga}(\pi)]<0$ is
non-trivial SPT phase protected by $x$-translation symmetry.

\begin{figure}[tb] 
\centering 
\includegraphics[height=1.1in]{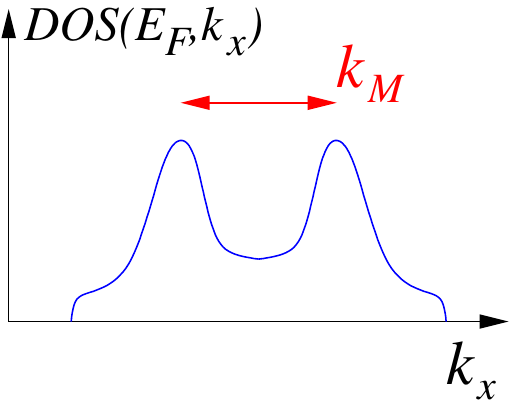} 
\caption{
The density of states per unit energy and per unit $\del k_x$.
}
\label{DOS} 
\end{figure}

\subsection{Examples of SMI-SC heterostructures}

In this section, we study various examples of  SMI-SC junctions, and identify
the conditions to realize the 1D TSC phase with weak MEC and a large gap.

\subsubsection{A SC bulk sample coupled to a spiral  magnetic stripe}

First let us consider a bulk $s$-wave superconductor on cubic lattice 
where $t_\mu$ in the model \eqref{HtDel3D} are given by
\begin{align}
 t_x=1,\ \ \  t_y=1,\ \ \  t_z=1.  
\end{align}
(In this paper, we always choose the nearest neighbor hopping in $x$-direction
$t_x=1$ and the lattice constant $a=1$.)  We choose the chemical potential to
be $\mu=-4$ so that the superconductor has a typical spherical Fermi surface.  The bulk superconductor is
coupled to a spiral  magnetic insulator stripe in $x$-direction on its $x$-$y$
surface (see Fig.  \ref{SCbMI}).  We take the superconducting gap to be
$\Del=0.05$ and the spiral wave vector $\v k^M=(\frac{2\pi}{3a},0)$.
The width of the spiral  magnetic insulator stripe is $d_y=a$ and $d_y=20a$
(the later is about the SC coherent length).  We obtain a phase diagram (see
Fig.  \ref{SCbSMI}) of the 1D SMI-SC junction, as we vary the MEC $J$.

From Figs. \ref{SCbSMI},
we see that the 1D TSC phase does
appear.  But the TSC phase only appears for very strong MEC $J>t_x$.  The
superconducting gap is only $\Del=0.05$.  Naively, we would expect
to obtain the 1D TSC phase when $J\sim \Del$.  

\begin{figure}[tb] 
\centering 
\includegraphics[height=1.5in]{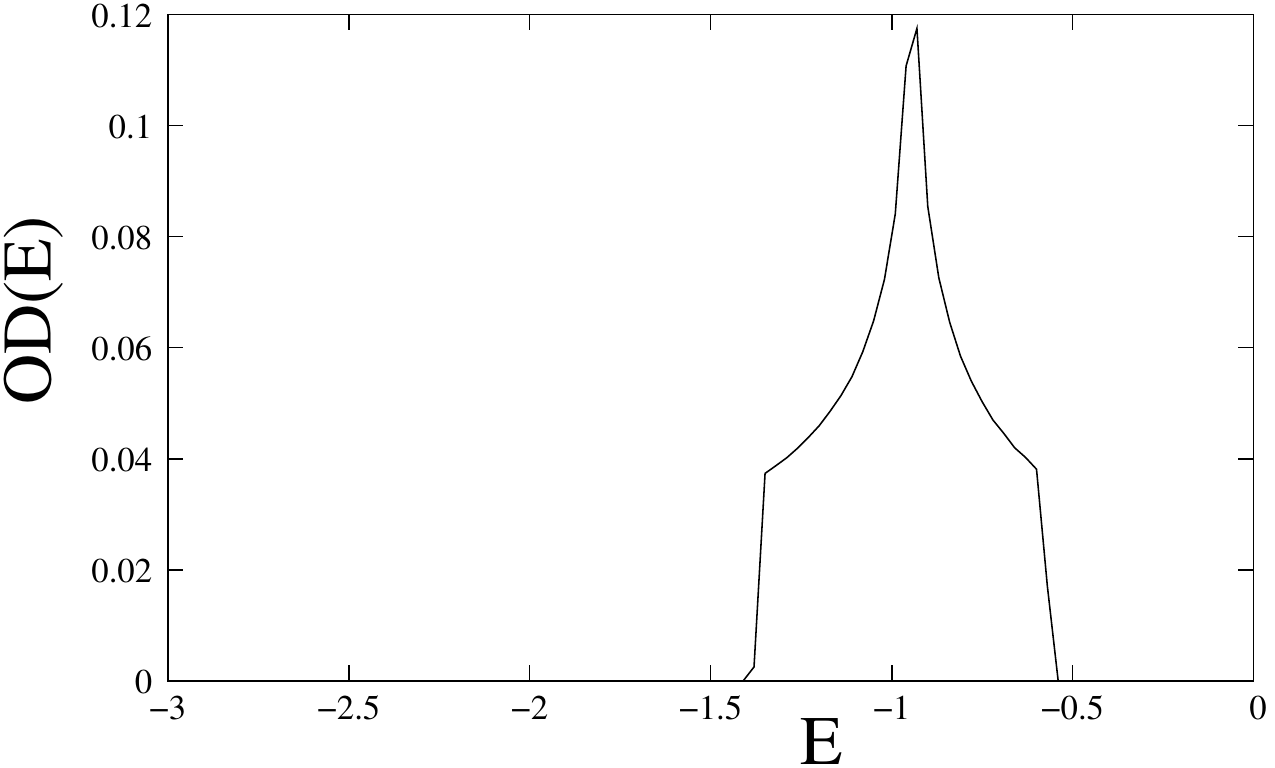} 
\caption{
The overlap density of states $OD(E)$ for our model of quasi 1D superconductor with
$t_x=1$, $t_y=t_z=0.2$ and $k^M=\frac{2\pi}{3a}$.
}
\label{DOSOverlapLz1tyz20_20SO3} 
\end{figure}

\begin{figure}[tb] 
\centering 
\includegraphics[height=1.5in]{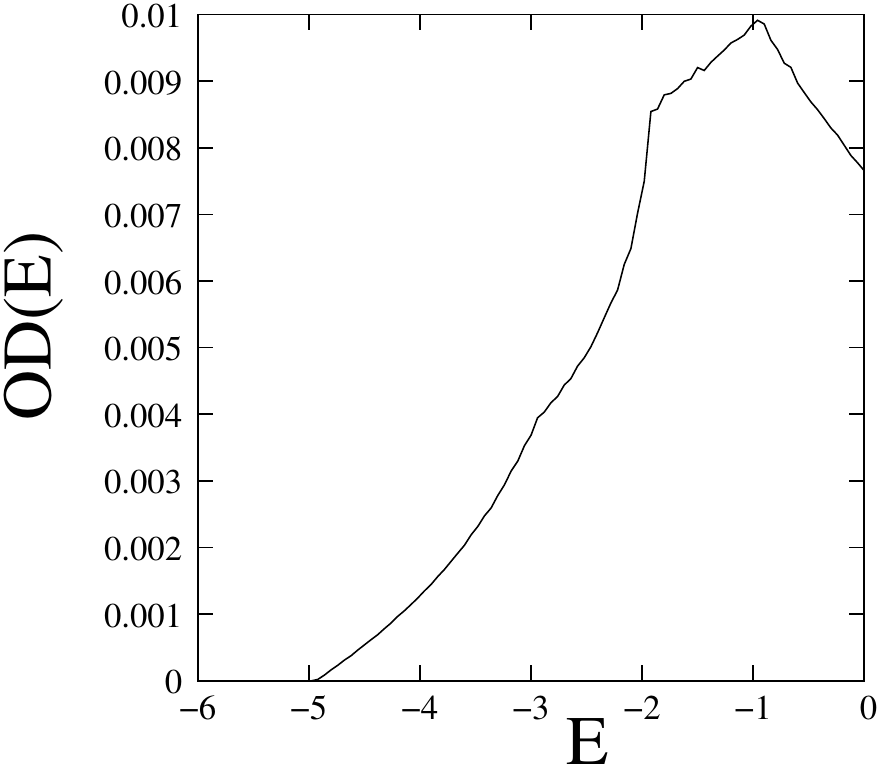} 
\caption{
The overlap density of states $OD(E)$ for typical superconductor with $t_x=t_y=t_z=1$ and
$k^M=\frac{2\pi}{3a}$, which is 10 times smaller than that of quasi 1D superconductor in
Fig. \ref{DOSOverlapLz1tyz20_20SO3}.
}
\label{DOSOverlapBulkSO3} 
\end{figure}

However, the magnetic insulator only interact with the first surface layer of
SC sample.  The energy shift $\Del E$ 
of a spin polarized level 
caused by the MEC is given by
\begin{align}
 \Del E \sim -\frac{J a}{l_\perp}+\frac{\hbar v_\perp}{l_\perp }
\end{align} 
where $a$ is the lattice constant, $l_\perp$ is the size of the wave function
of the  spin polarized level in the direction perpendicular to the 1D SMI-SC
junction, and the $v_\perp$ is the typical Fermi velocity in the direction
perpendicular to the 1D SMI-SC junction.  We see that MEC need to be $J >
\frac{\hbar v_\perp}{a}$ to push a spin polarized  levels  down to near zero
energy (\ie $\Del E \sim \Del$), so that those levels can organize into
the TSC phase.

Thus, the key to realize the 1D TSC phase at weaker MECs is to reduce transverse
velocity $v_\perp$.  This can be achieved if Fermi surface of the superconductor has flat
parts, and the 1D SMI-SC junction is perpendicular to the flat parts.

\begin{figure}[tb] 
\centering \includegraphics[scale=0.6]{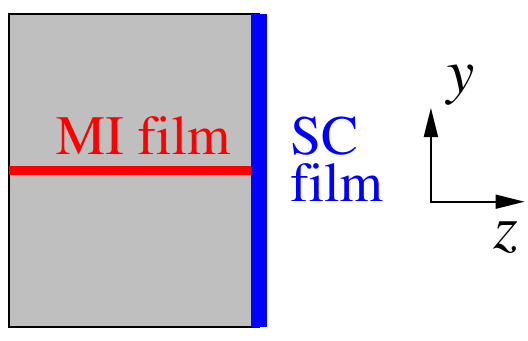} 
\caption{ 
A junction between a SC thin film and a magnetic insulator stripe on
the SC surface.  The SC surface is the $x$-$y$ plane and the magnetic stripe is
in $x$-direction.  
The  magnetic insulator stripe may also be replaced by a 1D
array of magnetic atoms that do not touch each other.
}
\label{SCfMI} 
\end{figure}

\begin{figure}[tb] 
\centering 
\includegraphics[height=1.2in]{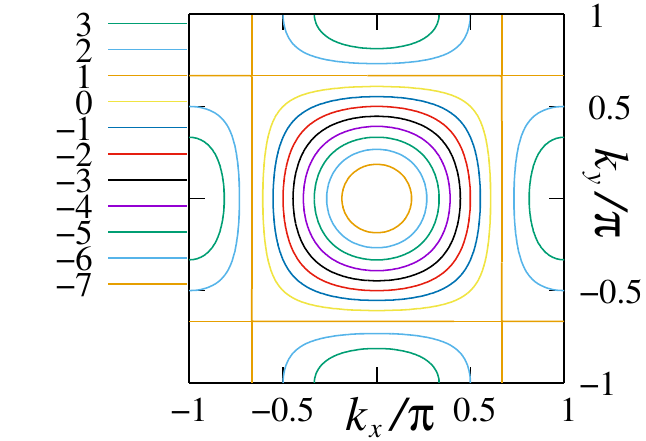} 
\caption{Parallel flat Fermi surfaces in monolayer superconductors.
}
\label{nestFS} 
\end{figure}

\begin{figure}[tb] 
\centering \includegraphics[scale=0.6]{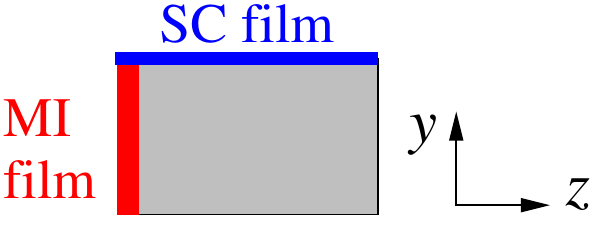} 
\caption{Another junction between a magnetic insulator thin film and a
$s$-wave SC thin film alone their edges.
} 
\label{SCMIa} 
\end{figure}

\begin{figure*}[tb] 
\centerline{
\includegraphics[height=2.2in]{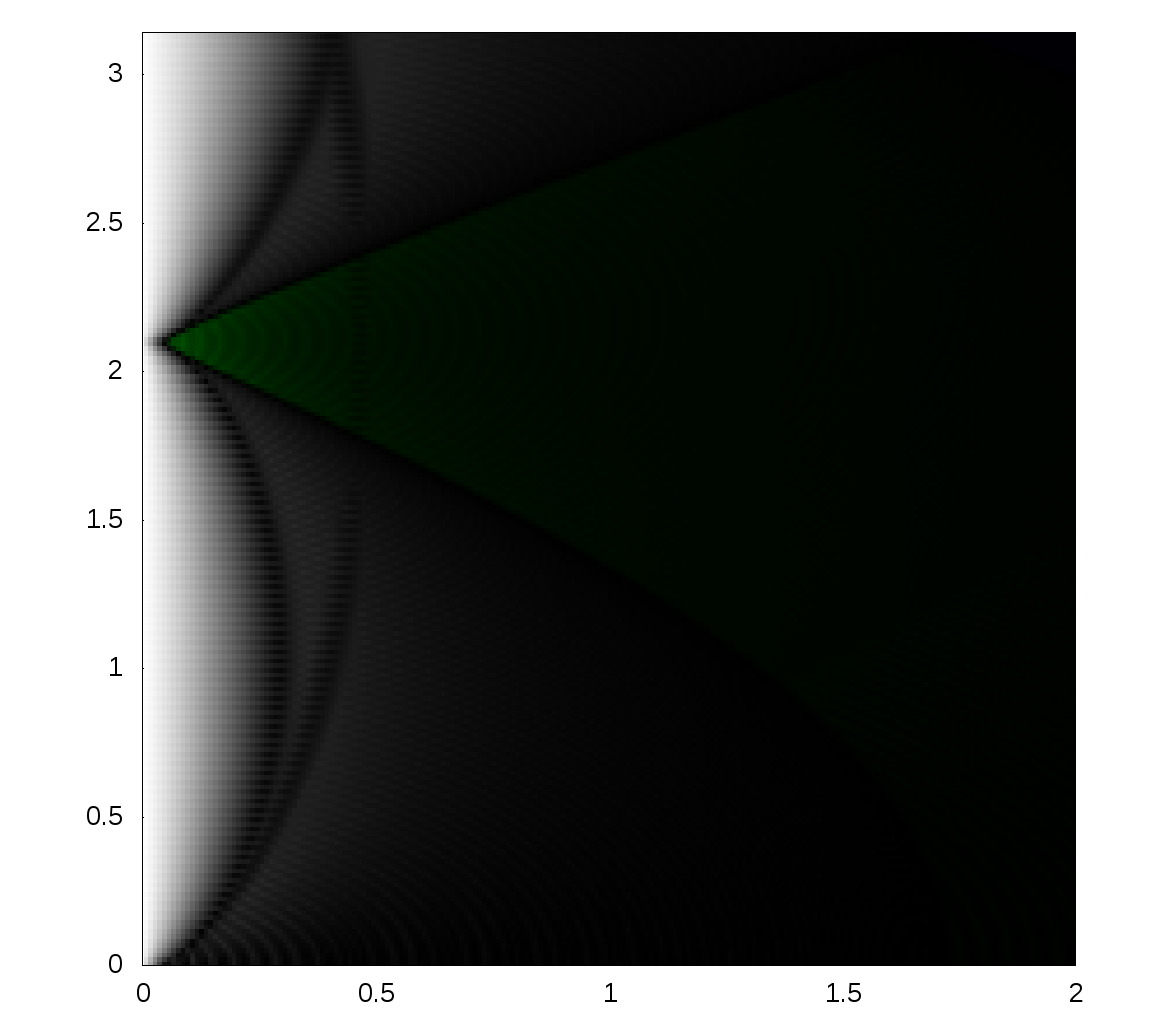} 
\includegraphics[height=2.2in]{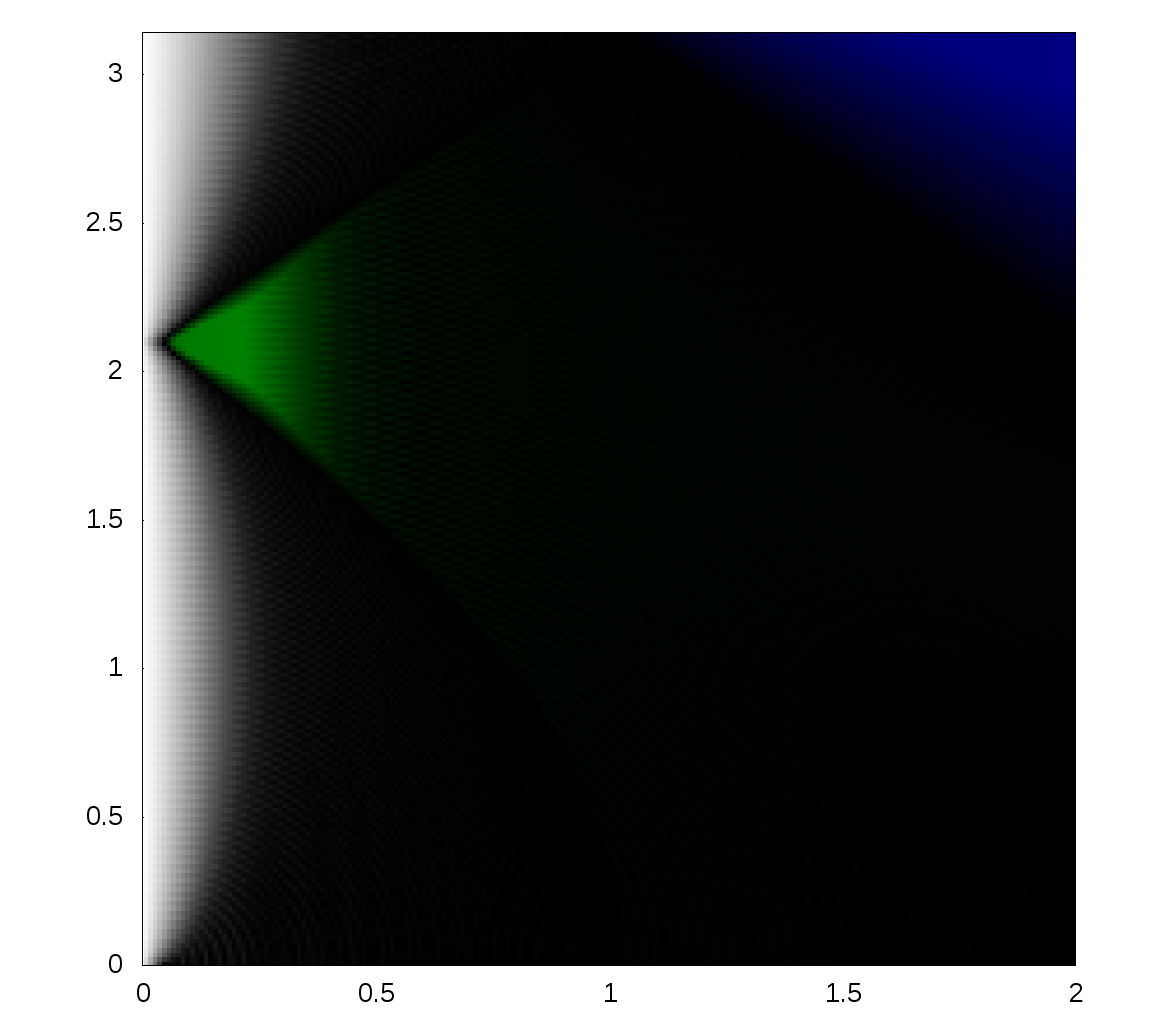} 
}
\caption{
The density plot of the lowest positive one-particle energy level $E_1$ for a
device made of flat-Fermi-surface SC monolayer film  (such as the NbSe$_2$
film).  The horizontal axis is the MEC $J$ and the vertical axis is the spiral
wave vector $k^M_x a$ where $k^M_y=0$.  The brightness is proportional to $E_1$
with maximum brightness corresponds to $E_1=\Del=0.05$.  Left: for device in
Fig. \ref{SCfMI} with an 1D array of magnetic atoms in the middle of SC film.
Right:  for device in Fig.  \ref{SCMIa} with an 1D array of magnetic atoms on
the edge of SC film.  
}
\label{BthN} 
\end{figure*}

\subsubsection{A quasi 1D SC thin film coupled to a spiral  magnet
along its edge}

One way to reduce $v_\perp$ is to simply use quasi 1D superconductors.
Recently, some quasi 1D superconductors were discovered, such as
Tl$_{2-x}$Mo$_6$Se$_6$ ($T_c=6.8$K, $x=0\sim 0.1$), K$_2$Cr$_3$As$_3$
($T_c=6.1$K), Rh$_2$Cr$_3$As$_3$ ($T_c=4.8$K), Cs$_2$Cr$_3$As$_3$ ($T_c=2.2$K),
RbCr$_3$As$_3$ ($T_c=7.1$K), KCr$_3$As$_3$
($T_c=5$K).\cite{BL11062405,BC14120067}  Without doping, the Fermi wave vector
along the 1D chain is $k_F =\frac{\pi}{2a}$ with $a=4.2 \AA$.

Let us assume that we can make a thin film of the quasi 1D superconductors, and we
consider SMI-SC junction in Fig. \ref{SCMI}, where the 1D chain is parallel to
the junction.  We choose the width of the magnetic stripe $d_y$ to be the same
as the thickness of the SC thin film.  We model the quasi 1D superconductors by reducing
$t_y=t_z$ from 1 to $0.2$, and obtain the phase diagram as in Fig.
\ref{Bmu1D1on1x40a}.  We see that the required MEC is much reduced to the scale
of the interchain coupling in the quasi 1D superconductor.  Even for small MEC, the gap of
the 1D TSC phase can be close to the superconducting gap $\Del$.
Increasing the thickness SC thin film from $d_y=a$ to $d_y=5a$ does not reduce
the 1D TSC gap by much.

However, the above-mentioned benefits come with a cost: the TSC phase can
appear at small MECs only for a window of chemical potentials $\mu$ (or a
window of the dopings of the SC film).  This is because, in order for the TSC
phase to appear at small MECs, the following conditions need to be satisfied:\\
(1) the density of states $D(E_F,k_x)$ per unit cell per unit $\del E$ and per
unit $\del k_x/2\pi$ has peaks, and\\ (2) the spiral wave vector $\v k^M$ can
connect the peaks (see Fig. \ref{DOS}).\\ As we change $\mu=E_F$, the
separation between the peaks can change, and we obtain TSC when the peak
separation matches $\v k^M$.  

To see the above condition more clearly, we introduce the following overlap of
the $k_x$ resolved density of states $D(E_F,k_x)$:
\begin{align}
 OD(E) = \int \frac{\dd k_x}{2\pi} D(E,k_x) D(E,k_x+k^M).
\end{align}
We plotted $OD(E)$ for our model of quasi 1D superconductor in Fig.
\ref{DOSOverlapLz1tyz20_20SO3}.  The overlap-density of states $OD(E)$ becomes
large ($\sim 0.05$) in a window $-1.5 < E < -0.5$ which match the window of the
chemical potentials, for the TSC phase to appear at small MECs (see Fig. \ref{Bmu1D1on1x40a}).

This is a key result in this section: To realize TSC phase for small MECs, we
need to find superconductor whose band structure gives rise to large $OD(E)$, such as
$OD(E) > 0.04$ 
in $t_x =1$ unit as one can see from Fig. \ref{DOSOverlapLz1tyz20_20SO3}.  For
a typical superconductor with $t_x=t_y=t_z=1$, the overlap-density of states $OD(E)$ is
given by Fig.  \ref{DOSOverlapBulkSO3}, where $OD(E) < 0.01$ for all $E$.  At
chemical potential $E=-4$, $OD(-4) \sim 0.001$.  This explains why we need a
very large MEC $J > 1.5 $ to realize the topological SC phase.

The phase diagram Fig.  \ref{Bmu1D1on1x40a}  is for a very large spiral wave
vector $k^M =\frac{2\pi}{3a}$.  Such a large spiral vector can appear in
triangle-lattice magnetic materials, such as CuCrO$_2$.  We also see that the
TSC phase does not appear for undoped quasi 1D superconductors with chemical potential
$\mu=0$ and Fermi wavevector $k_F=\frac{\pi}{2a}$.  To obtain TSC, we must dope
the quasi 1D superconductors to shift chemical potential $\mu$ away from zero.  The 10\%
doping of the quasi 1D superconductors has been achieved, which corresponds to $\mu=\pm
0.3$.  For such doping, the TSC phase does not appear for small MEC if the
thickness of the SC thin film is $d_y=a,2a$. But for $d_y=5a$, the topological
SC phase appear for $\mu=0.3$ and $J=0.5$.  However, it is highly
non-trivial to find a match between quasi 1D superconductor and SMI so that they can form a
good junction as in Fig. \ref{SCMI}.

\subsubsection{Superconductors with flat Fermi surfaces}

\begin{figure}[tb] 
\centering 
\includegraphics[height=1.3in]{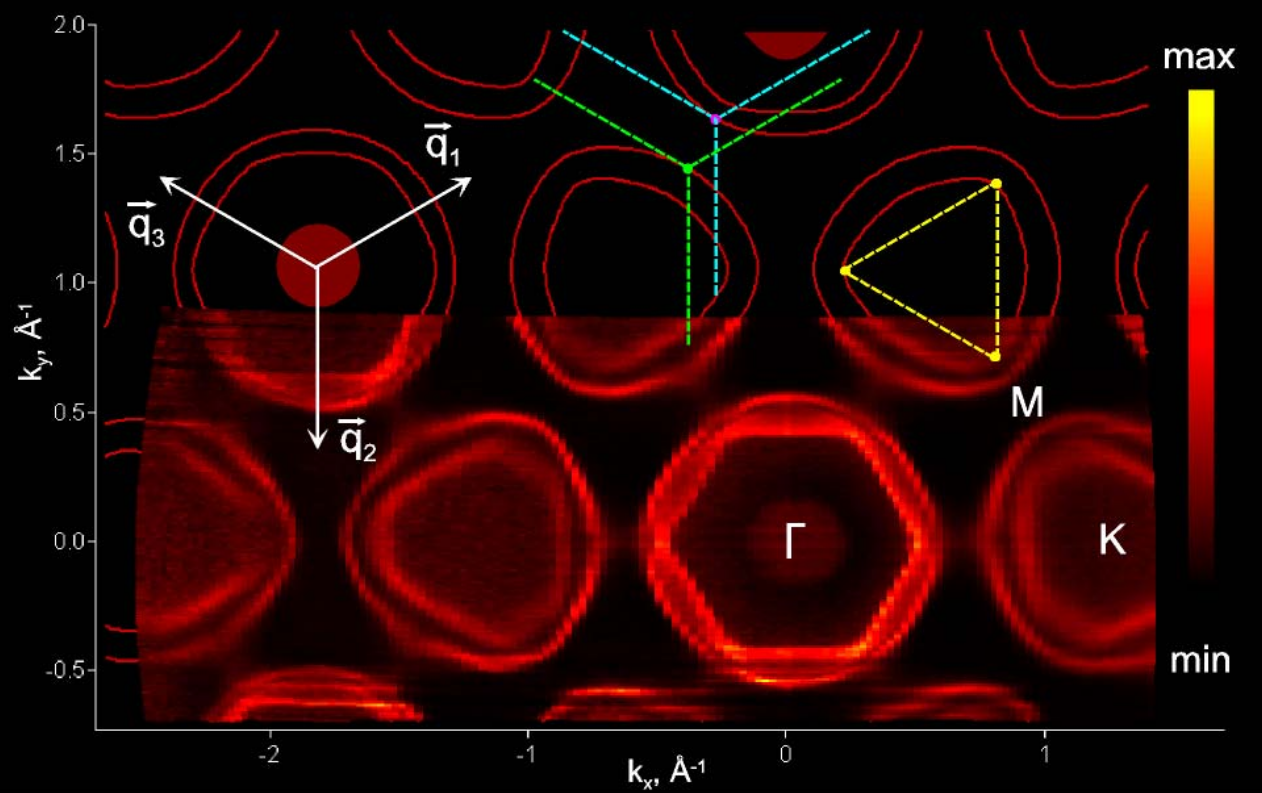} 
\caption{
The Fermi surfaces of NbSe$_2$ contains tube-like parts and parallel
sheets. NbSe$_2$ is superconducting with $T_c=7.2$K
(see \Ref{PhysRevLett.102.166402} and \Ref{xi2016gate}).
We can use a monolayer NbSe$_2$ film to make a device in
Fig. \ref{SCdotF}.
A model phase diagram for such a device is given in Fig. \ref{BthN}.
}
\label{NbSe2} 
\end{figure}
\begin{figure}[tb] 
\centering 
\includegraphics[height=1.3in]{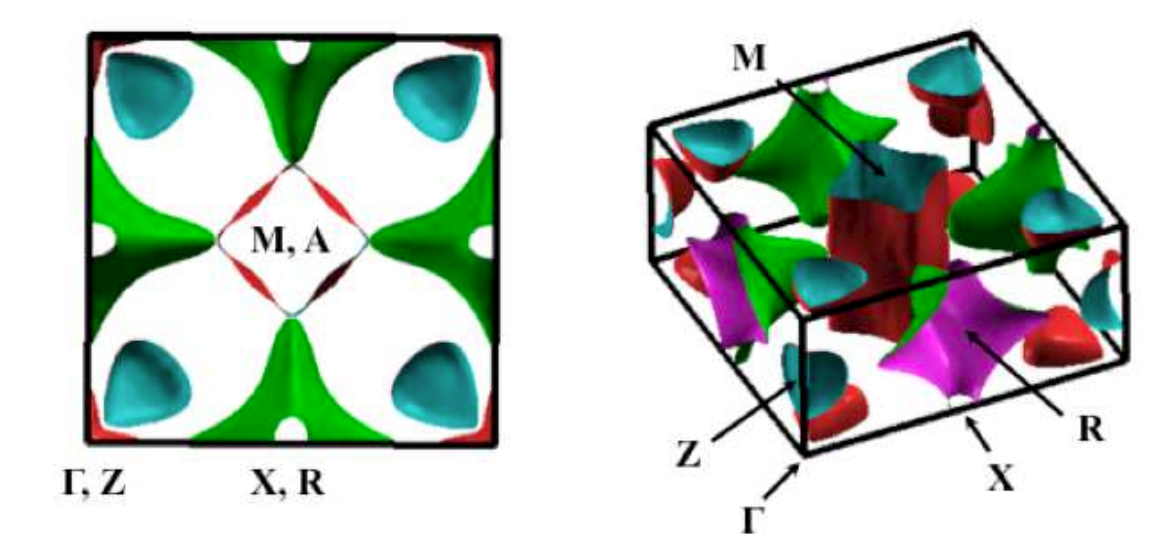} 
\caption{
The Fermi surfaces of BaTi$_2$A$_2$O contains tube-like parts and parallel
sheets.  Here A = As, Sb, Bi.  For Ba$_{1-x}$K$_x$Ti$_2$Sb$_2$O, the $T_c$ is
$1.2$K at $x=0$ and $6.1$K at $x=0.12$  (see \Ref{NM160202024}).
}
\label{BaTi2A2O} 
\end{figure}

\begin{figure}[tb] 
\centering 
\includegraphics[scale=0.4]{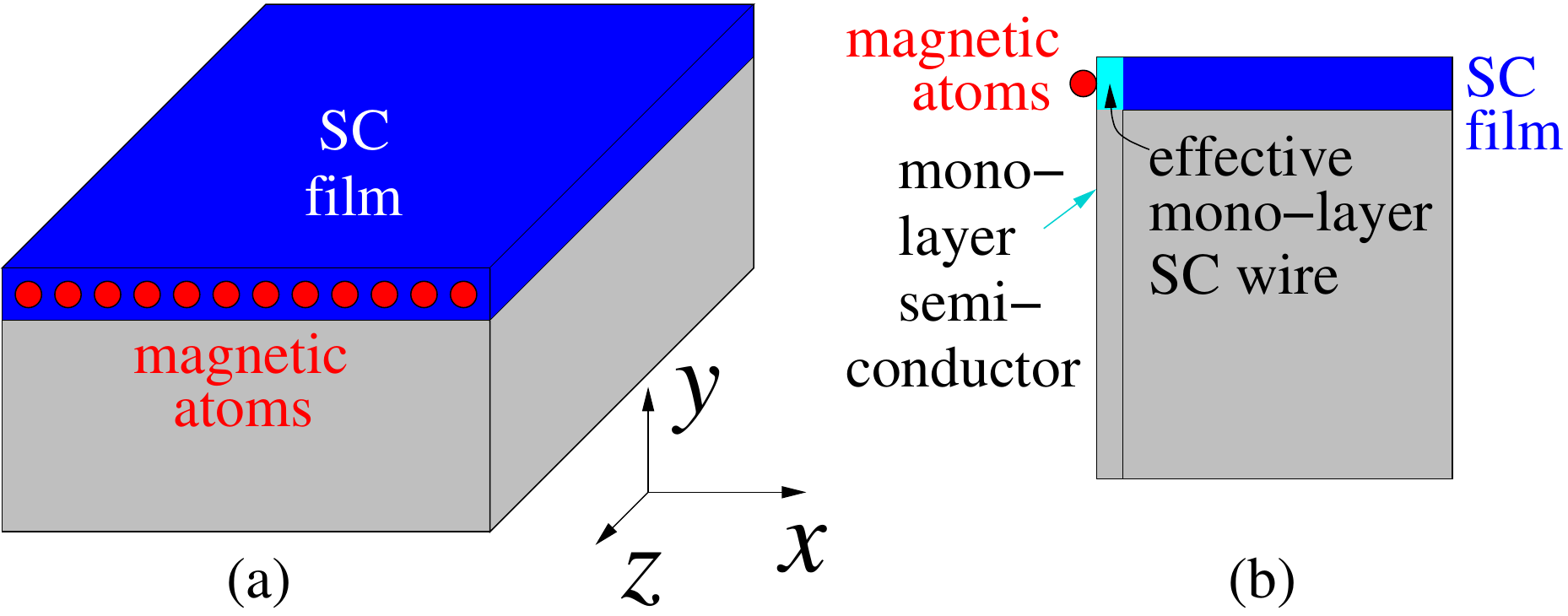}
\caption{ 
Two designs to engineer an SC wire.  (a) If the Fermi surface contains a
tube-like part, the axis of the tube should be in $z$-direction.  If the Fermi
surface contains a pair of parallel sheets, the sheets should be in $yz$-plane.
In this case, the edge will behave like a SC wire.  (b) For a generic SC, we
may evaporate a monolayer semiconductor on the cleave edge of the SC film to
engineer an effective monolayer SC wire on the edge of the SC film.
}
\label{SCdotF} 
\end{figure}
\begin{figure*}[tb] 
\centering 
\includegraphics[height=1.8in]{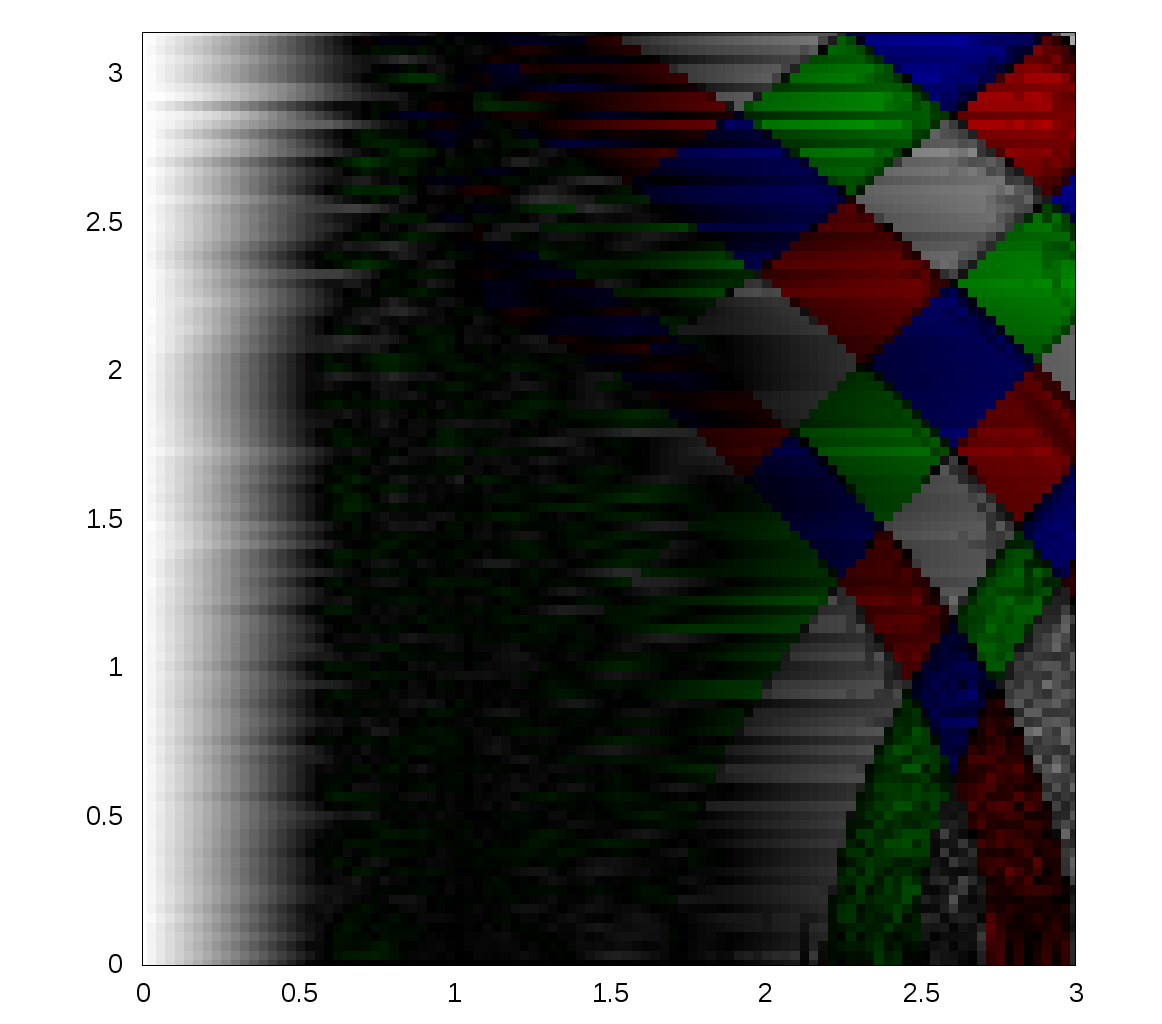} 
\includegraphics[height=1.8in]{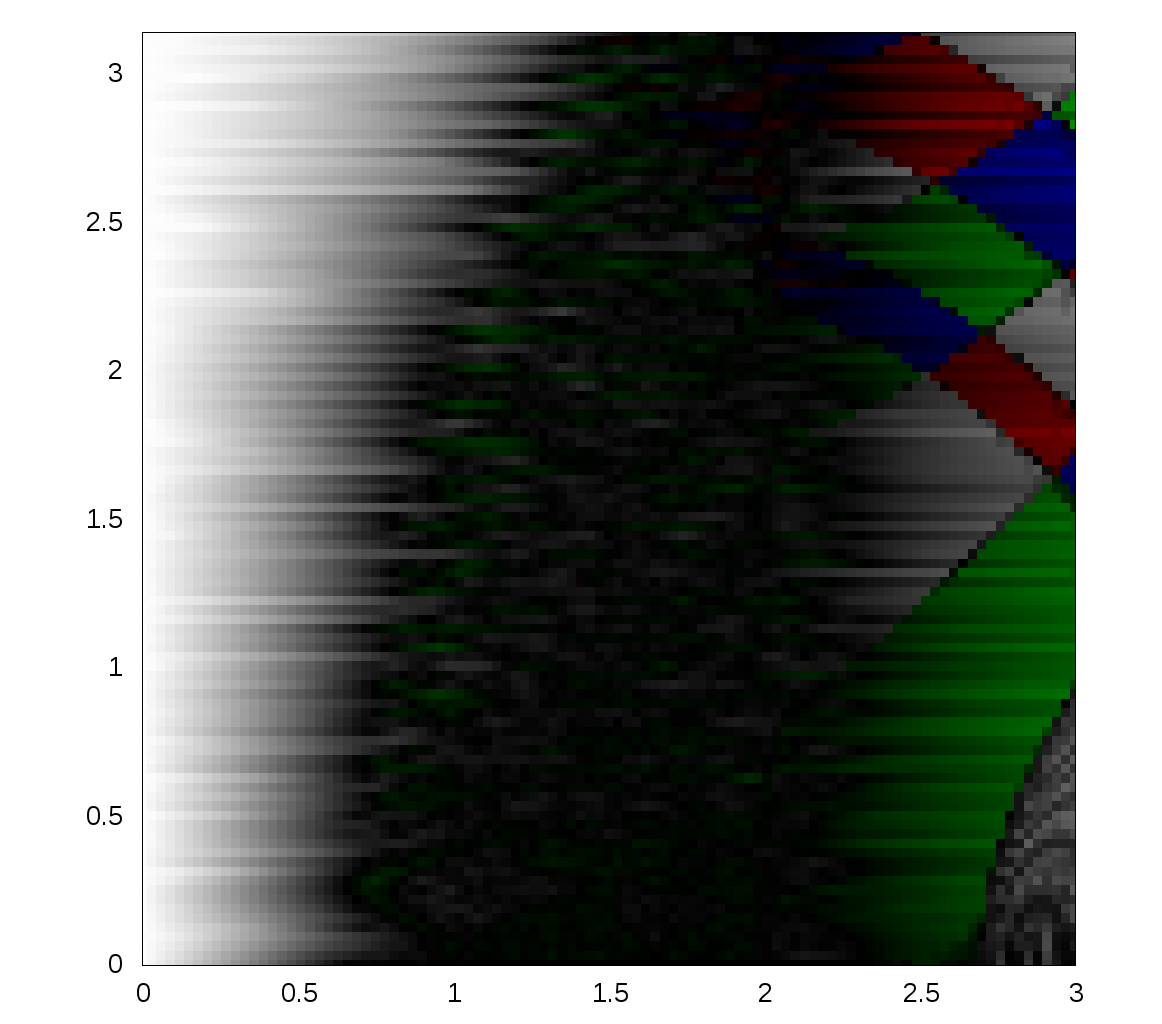} 
\includegraphics[height=1.8in]{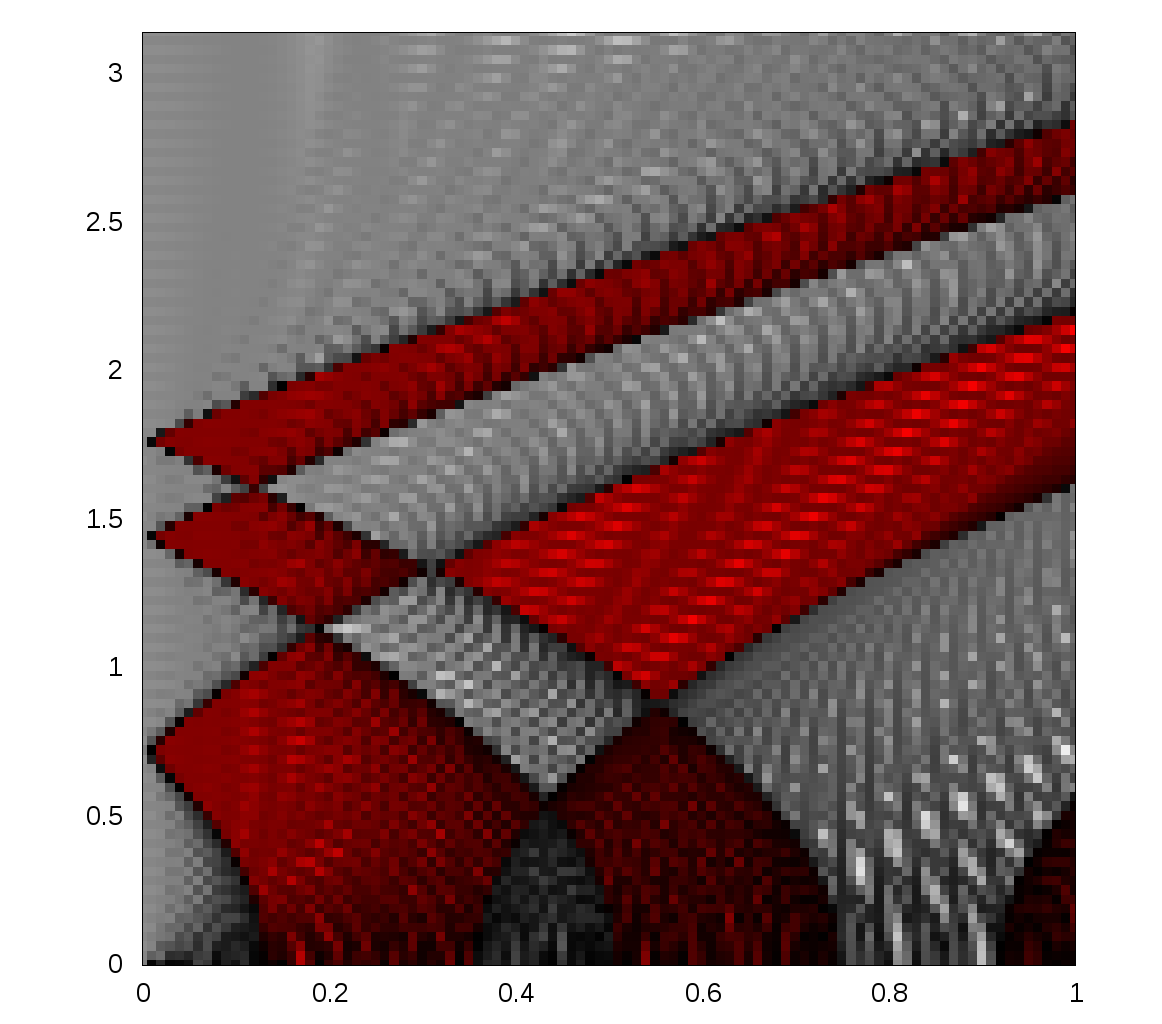} 
\caption{ 
The phase diagram of model \eq{HtDel3D}.  The horizontal axis is the MEC $J$
and the vertical axis is the spiral wave vector $k^M_x a$.  Left: no monolayer
on the cleave edge of the SC thin film.  Middle: a monolayer of ``similar''
atoms ($\mu_S=-4$).  Right: a monolayer of ``very different''  atoms
($\mu_S=3$).  The red and green areas represent the TSC phase (with
$\text{Pf}[M_\ga(0)]<0, \text{Pf}[M_\ga(\pi)]>0$ for red and
$\text{Pf}[M_\ga(0)]>0, \text{Pf}[M_\ga(\pi)]<0$ for green, see 
Supplementary Materials
Section III).  The grey area represents the trivial gapped phase (with
$\text{Pf}[M_\ga(0)]>0, \text{Pf}[M_\ga(\pi)]>0$).  The blue area represents
the SPT phase protected by translation symmetry (with $\text{Pf}[M_\ga(0)]<0,
\text{Pf}[M_\ga(\pi)]<0$).  The brightness represent the energy gap of the
corresponding 1D state.  The full brightness corresponds to a gap value
$\Del=0.02$.
}
\label{thBkO} 
\end{figure*}

We have seen that a small transverse velocity is very helpful to obtain 1D TSC
phase for small MECs.  In addition to quasi 1D superconductors, another way to
realize small transverse velocity is to use superconductor with flat Fermi
surfaces, such as NbSe$_2$ \cite{PhysRevLett.102.166402} (see Fig.
\ref{NbSe2}) and BaTi$_2$A$_2$O where A = As, Sb, Bi (see Fig.
\ref{BaTi2A2O}). In particular, NbSe$_2$ has two advantages. First, it is a
layered material, with two-dimensional character. It can even be made into a
monolayer. Second, NbSe$_2$ can be doped and its chemical potential can be
tuned continuously.\cite{xi2016gate}  

We construct the device by putting magnetic atoms on the above superconductors.
The magnetic atoms form a 1D array along the normal direction of the flat Fermi
surface.  It is important that the magnetic atoms do not touch each other, so
that the magnetic atoms do not form a 1D metallic wire.

We can model such superconductors and their flat Fermi surfaces by choosing 
\begin{align}
 t_x=t_y=t_{x+y}=t_{x-y}=1, \ \ t_z=0,\ \
\v k^M=(k_x^M,0,0).
\end{align}
(We only consider monolayer SC thin film by choosing $t_z=0$.)
When the Fermi energy is near $E_F=1$, there are two pairs of parallel flat
Fermi surfaces, which are separated by $\frac{2\pi}{3a}$ (see Fig.
\ref{nestFS}).  Putting magnetic atoms with a spiral magnetic moment on such a
SC,  we obtain the phase diagram as in Fig. \ref{BthN}.  We see that 1D TSC
phase appears for very small MECs.  If we put the array of magnetic atoms on
the edge of the SC thin film, the maximum gap of the TSC is about
$0.5\Del$.  If we put the array of magnetic atoms in the middle of the SC
thin film, the maximum gap of the TSC is about $0.3\Del$.

\subsection{Engineer a superconducting wire}
\label{cleave}

If the Fermi surface of SC material contains tube-like parts or parallel sheets
(see Fig. \ref{NbSe2} and Fig. \ref{BaTi2A2O}), then the SC material is
effectively a quasi-2D or a quasi-1D material for our purpose of making 1D TSC.
In this case, we can use a SC thin film to mimic a quantum wire.  Instead of
putting isolated magnetic atoms on a quantum wire, we can put the isolated
magnetic atoms on the side of the SC thin film (see Fig.  \ref{SCdotF}a). The
thickness of the SC film should be a few times $2\pi/k_F$.  We expect that a
spiral magnetic order on the magnetic atoms will develop spontaneously to
produce a 1D $p$-wave topological superconductor with MZM at ends of the
chain.

For generic SC materials without tube-like Fermi surfaces or parallel Fermi
surfaces, we need a  very strong MEC $J > t_x$ to realize 1D TSC (see Supplementary Materials Fig.
\ref{thBkO}:Left).  However, if the superconductor has well-localized surface
states, those surface states on the edge of the SC film may couple strongly to
the magnetic atoms. In this case, we may not need a strong MEC to realize 1D
TSC and its MZM.

We may even engineer those surface states by evaporating a thin layer of
certain atoms on the edge of the film.  For example, we can evaporate a
monolayer of semiconductor on the cleaved edge of the SC thin film.  We hope
the SC thin film to dope the semiconducting monolayer, and the semiconducting
monolayer becomes superconducting due to the proximity effect.  Note that the
superconducting electrons in the semiconducting monolayer are well localized
inside the monolayer.  This way, we obtain an effective monolayer SC wire on
the cleaved edge of the SC film, which is well modeled by \eqn{HtDel}.  If we
can find a superconducting/metallic atom that only sticks to the SC thin film,
rather than the substrate, then we can evaporate a monolayer of such
superconducting/metallic atoms to engineer an effective monolayer SC wire on
the edge of the SC film (See Fig. \ref{SCdotF}b).

To confirm the validity of the above design, we consider the following model to
describe the SC thin film
\begin{align}
\label{HtDel3D}
 H &=\sum_{\v i,\v \mu} 
[-t_{\v \mu} c_{\v i+\v\mu,\al}^\dag c_{\v i,\al} +h.c.]
+ \sum_{\v i} \v B_{\v i}  c_{\v i}^\dag \v \si c_{\v i} 
\nonumber\\
& 
+\sum_{\v i} 
[\Del (c_{\v i \up} c_{\v i\down}-c_{\v i \down} c_{\v i\up}) +h.c.]
-\sum_{\v i} \mu c_{\v i}^\dagger c_{\v i},
\end{align}
where $\v \mu =\v x,\v y, \v z$, 
\begin{align}
\v B_{\v i}  = B_{i_z} \Big(\sin(k^M_x i_x a),0,\cos(k^M_x i_x a)\Big)
\end{align}
describes the MEC to the magnetic atoms.  Here
\begin{align}
 B_{i_z} = J \del_{i_z,0},
\end{align}
\ie the MEC only couples to the first layer (marked by $i_z=0$) on the SC
surface.  We choose $t_x=t_y=t_z=1$, $\Del = 0.02$, $\mu=-3$, and
$(L_x,L_y,L_z)=(200,10,60)a$.  Such a SC film has an electron-like Fermi
surface and a superconducting coherent length $\xi \sim 30a$, We also assume
one magnetic atom per site (\ie no randomness), which gives us a phase diagram
Fig.  \ref{thBkO}:Left.  We see that only for MEC $J > 1.5$ can the TSC phase
appear.

We can model the monolayer of semiconductor by changing the chemical potential
to $\mu_S$ at the SC surface layer (at $i_z=0$).  We also set the proximity
induced superconducting gap at the surface layer $i_z=0$ to be
$\Del_S=0.01=0.5\Del$.  If we choose $\mu_S=-4$, the surface monolayer will
have a smaller electron-like Fermi surface.  We obtain a phase diagram Fig.
\ref{thBkO}:Middle with TSC phase for $J > 2.2$.  We see that adding a
monolayer of ``similar'' atoms does not help. 

If we choose $\mu_S=3$, the surface monolayer will have a hole-like Fermi
surface which is very different from the bulk SC film.  We obtain a phase
diagram Fig.  \ref{thBkO}:Right.  We see that adding a monolayer of ``very
different'' atoms makes the TSC appear at very weak MEC $J \sim \Del_S$.
For MEC $J=0.05$, the gap of TSC is about $\Del_S=0.01$.

Thus it is important that the monolayer is formed with atoms that have a very
different band structure than the bulk SC film, such as one has electron-like
Fermi surfaces and the other has hole-like Fermi surfaces.  We see that the
cleave-edge overgrowth of a monolayer of ``very different'' atoms  on a SC
thin film, plus a dilute layer of isolated magnetic atoms, is a practical way
to realize MZM.

\bibliography{../../../bib/wencross,../../../bib/all,../../../bib/publst,MZM} 

\end{document}